\documentclass[10pt]{article}
\usepackage[utf8]{inputenc}
\usepackage{cite}
\usepackage{mathtools}
\usepackage{longtable}
\usepackage{multirow}
\usepackage{amsmath}
\usepackage{graphicx}
\usepackage{amsfonts}
\usepackage[margin=.4in]{geometry}
\usepackage{adjustbox}
\usepackage{caption}
\usepackage{subcaption}
\title{\bf   Shadows and Strong Gravitational Lensing by Van der Waals Black Hole in Homogeneous Plasma }
\author{\bf Niyaz Uddin Molla\footnote{e-mail:{niyazuddin182@gmail.com}}~ and Ujjal Debnath\footnote{e-mail: {ujjaldebnath@gmail.com}}\\
Department of Mathematics, Indian Institute of Engineering Science\\ and Technology, Shibpur, Howrah-711103,India.}

\begin{document}
\maketitle

\begin{abstract}

In this paper, we first analyze the horizon structure of the Van
der Waals(VdW) black hole and then investigate its shadow in the
absence of a plasma medium as well as the presence of a
homogeneous plasma medium. We find that both the Van der Waals
parameters $a$ and $b$ have a significant effect on the shadow of
the black hole. We also observe that the radius of the shadow in a
homogeneous plasma medium decreases while parameter $\sigma
=\frac{\omega_p}{\omega_{\infty}}$ ( the ratio of plasma frequency
and photon frequency) increases and the radius of the shadow
inhomogeneous plasma medium is larger than the vacuum medium. We
also discuss the strong gravitational lensing in a homogeneous
plasma medium. We observe that the photon sphere radius,
deflection limit coefficients and deflection angle in the strong
field are highly affected by the presence of a homogeneous plasma
medium. We also find that the deflection angle in the strong field
limit by the Van der Waals black hole with the homogeneous plasma
is greater than that of the Vacuum medium. Further, we discuss the
observables quantities angular position $\theta_{\infty}$,
separation $S$ and magnification $r_{mag}$ by taking the example
of a supermassive black hole in the strong field limit with the
effects of homogeneous plasma. It is concluded that the Van der
Waals parameters $a$, $b$  and homogeneous plasma medium have a
significant effect on both the shadows and strong gravitational
lensing.

\end{abstract}

\section{Introduction}
One of the most powerful and important tools in astrophysics as
well as cosmology is gravitational lensing(GL), which deals with the
deflection light rays passing through the gravitational field. It
has been successfully employed to explain for probing the strong
characterization of gravity. Accordingly, the deflection angle of
photon rays, the GL can be divided into
scenarios, one of them  is weak GL when the
deflection angle is small, and another one is strong GL, when the deflection angle of photon rays becomes so
larger. Strong GL was first investigated by
darwin in 1959 \cite{D.1.} that the photons rays passing near to a
black hole(BH) may have a large deflection angle and could make multiple loops around the BH before reaching to the observer.
Leter, Virbhadra and  Ellis \cite{V.E.}, and Frittelli et al.
\cite{F.K.N.} derived the exact lens equation regardless of
background spacetime for arbitrary large value of deflection
angle. After that Bozza et.al \cite{B.C.I.S.} developed an useful
method to  obtain the deflection angle of light rays  by a compact
object in the strong field limit, and they found  the logarithmic
divergence of the deflection  angle  for Schwarzschild  BH
in the strong field region. Later, Bozza et al.\cite{B.V.1.}
extended the previous analysis  for any general  static,
asymptotically spherically symmetric spacetime. In the last
decades, strong GL regained more
attention.\cite{A.1.,L.,C.,N.,S.I.,M.U.1.}.

In 2019, the first image of M87* was observed by the Event horizon
Telescope (EHT)  \cite{A.1.,A.2.,A.3.,A.4.,A.5.,A.6.}, which gives
us the deeper understanding of BH physics and testing the
various types of black BHs. One of the most important features
of this image is that the event horizon of the BH is
surrounded by a dark region called BH shadow and a
bright-light ring encircled around the BH shadow. It is
also known from some literature
\cite{F.M.A.,A.E.,T.L.B.,A.A.A.}that the BH shadow is
created by the BH effect.The study of
BH shadow has been investigated by many authors for the
different spacetimes \cite{ J.,dv.,A.E.G.,A.E.1.,
Y.N.C.S.,A.A.A.G.,A.G.1.,T.L.B.,H.M.,B.Y.,G.,
T.,W.L.,A.A.K.A.C.,A.E.2.,B.F.,A.A.A.,W.C.J.,S.S.,G.P.L.,S.,K.S.G.,
K.S.A.G.,K.G.,N.H.A.A.,W.Z.L.M.,P.A.,A.P.J.,C.E.J.M.,C.G.,N.T.W.
}.

The BH as a thermodynamic system which is almost
equivalent to the classical thermodynamic system. For the BH in Ads spacetime, the Ads charged BH is almost equivalent
to the Van der Waals  (VdW)fluid system. The BH is not only
treated as a thermodynamic system, but it's also can be treated as
a gravitational system in our universe.

The thermodynamic properties of the BH have an important
role in the quantum gravity theory. Due to the Ads /CFT
corresponds, the physics of asymptotically Ads BH great
attention in the last decades.
 In the point of veiw of extended phase space \cite{K.M.,G.M.K.} where the cosmological constant ($\Lambda<0$) is behave as a thermodynamic pressure  $p$ \cite{C.M.,C.C.K.}
 \begin{equation}\label{1}
  p_{\Lambda}=-\frac{\Lambda}{8\pi}=\frac{3}{8\pi l^2}
  \end{equation}

is allowed to vary in the first law of thermodynamic
\begin{equation}\label{2}
\delta M= T\delta S+V\delta p+.....
\end{equation}
where the thermodynamic volume  V,thermodynamically conjugate to p is given by
\begin{equation}\label{3}
V=\frac{\partial M}{\partial T}_{S,...}
\end{equation}
Here $M,~ S,~ T$ are the mass, entropy, temperature of the BH respectively.

In \cite{00.R.K.M.} Rajagopal et al. have shown asymptotatically
BH metric  whose thermodynamic matches exactly with the
Van der Waals(VdW) BH. The VdW fluid described by the
closed form two parameter equation of state:
\begin{equation}\label{4}
T=(p+\frac{a}{v^2})(v-b)
\end{equation}

Here, the specific volume of the VdW fluid is given by
$v=\frac{V}{N}$, Where N is the degrees of freedom of the fluid
and V is the volume occupied by the fluid. The constant $a>0$
measures the intermolecular forces in the fluid and the constant
$b>0$ measures the volume of the molecules.

In recent years,Van der Waals  BH has been investigated
in various astronomical aspects such as, BHs as heat
Engine \cite{R.U.}. In this work, we discussed the shadow and
strong GL with the effects of homogeneous
plasma.

A Plasma is a dispersive medium, and when light rays pass through
the dispersive medium, they are refracted by this medium before
reaching to the observer.  From an astrophysical point of view,
all photons mostly go through a plasma medium.  On the other hand,
the plasma medium can affect the angular position of an equivalent
image, giving various wavelengths in observation.

This is the most intriguing and important reason why one needs to
consider the plasma medium in the analysis of GL.  To calculate the deflection angle  of the photon rays
and shadows of the BH in the presence of  plasma  and
gravity, we apply the  method \cite{P.T.0.,T.3.} which has been
discussed in detail in the books (see \cite{P.1.,S.1.}).

The effect of plasma on  the shadow of BHs and wormholes
have been widely  investigated by some authors
\cite{B.T.4.,A.T.S.A.,P.T.4,A.J.A.S.,H.D.L.}.GL
by the BH in a homogeneous plasma, the medium was studied
in \cite{B.T.1.,B.T.2.,M.A.T.,E.M.,A.A.,R.,P.T.B.,T.B.,L.D.J.}.

The paper is arranged as follows: In Sec.2, we review the VdW spacetime and the null geodesics equation. In Sec.3, we
study the shadows of the VdW BH in the absence
of a plasma medium. In Sec.4, We study the shadows of the Van der Waals  BH with the presence of plasma medium. We discuss
the strong GL of Van der Waals BH with
the effects of homogeneous plasma medium in Sec.5. Further, we
also discuss the strong observable quantity in Sec.6. Finally, we
discuss and conclude the study in Sec.7.

\section{Van der Waals spacetime and Null geodesic equations}
We start with a  spherically  symmetric, static  Ads spacetime, constructed by Rajagopal \cite{00.R.K.M.}  as

\begin{equation}\label{5}
ds^2=-f(r)dt^2+ \frac{dr^2}{f(r)} +r^2 d\theta^2 +r^2 \sin^2\theta d\phi^2
\end{equation}
  where
  \begin{equation}\label{6}
    f(r)=2\pi a-\frac{2M}{r}+  \frac{r^2}{l^2}(1+\frac{3b}{2r}) -\frac{3\pi a b^2}{r(2r+3b)} -\frac{4\pi a b}{r}log(\frac{r}{b} +\frac{3}{2})
    \end{equation}
    The event horizon of the BH locates at $r=r_{+}$.
  The horizon radius $r_{+}$ is determinants by the solutions of the
  equation $ f(r_{+}) = 0 $ i.e.,
   \begin{equation}\label{7}
  2\pi a-\frac{2M}{r_{+}}+  \frac{r_{+}^2}{l^2}(1+\frac{3b}{2r_{+}}) -\frac{3\pi a b^2}{r_{+}(2r_{+}+3b)} -\frac{4\pi a b}{r_{+}}log(\frac{r_{+}}{b} +\frac{3}{2}) =0
    \end{equation}
  whose real roots  are shown graphically in Fig.1 and Table.1

It has been shown that the spacetime solution (5) does not
satisfied any of the standard energy conditions everywhere out
sides the horizon but for the sufficiently small pressure $p$,
energy condition can be satisfied in the neighbouring of the BH horizon. The Lagrangian represents the motion of light rays
around the spacetime (\ref{5}) is  given by

\begin{equation}\label{8}
2\mathcal{L}=-f(r)\dot{t}^2+ (f(r))^{-1} \dot{r}^2 +r^2 \dot{\theta}^2 +r^2 \sin^2\theta \dot{\phi}^2
\end{equation}
 which is used to calculate the geodesic equations.
Here dot indicates  the derivative w.r.t  the affine parameter. The null geodesic equations for photon motion are given by

   \begin{equation}\label{9}
   \dot{t}=\frac{E}{f(r)}
       \end{equation}
   \begin{equation}\label{10}
   r^2\dot{\theta}=\pm\sqrt{\Theta}
   \end{equation}
   \begin{equation}\label{11}
   \dot{\phi}=\frac{L}{r^2 \sin^2\theta}
   \end{equation}
   \begin{equation}\label{12}
   r^2\dot{r}=\pm\sqrt{\mathcal{R}}
   \end{equation}

   where $$\Theta=K-\frac{L^2}{\tan^2\theta}$$
   $$ \mathcal{R} = E^2 r^4 -r^2 (L^2+k)f(r)$$
   and L ,E are the angular momentum and Energy of the photon respectively.
    From the equation (\ref{10}), one can write the equation as

    \begin{equation}\label{3}
    \dot{r}^2+ V_{eff} =0
\end{equation}
   where $V_{eff}$ is the effective potential function  and it is obtained by $$V_{eff}= \frac{f(r)}{r^2}(K+L^2)-E^2$$

Now, we change the effective potential function $V_{eff}$ with respect to the new impact parameters such as $\xi=\frac{L}{E}$  and $\eta=\frac{K}{L^2}$
\begin{equation}\label{14}
V^{\prime}_{eff}= \frac{E^2}{r^2}\{(\eta+\xi^2)f(r)-r^2\}
\end{equation}

  \begin{figure}
 \begin{subfigure}[b]{.4\textwidth}
    \includegraphics[width=85mm]{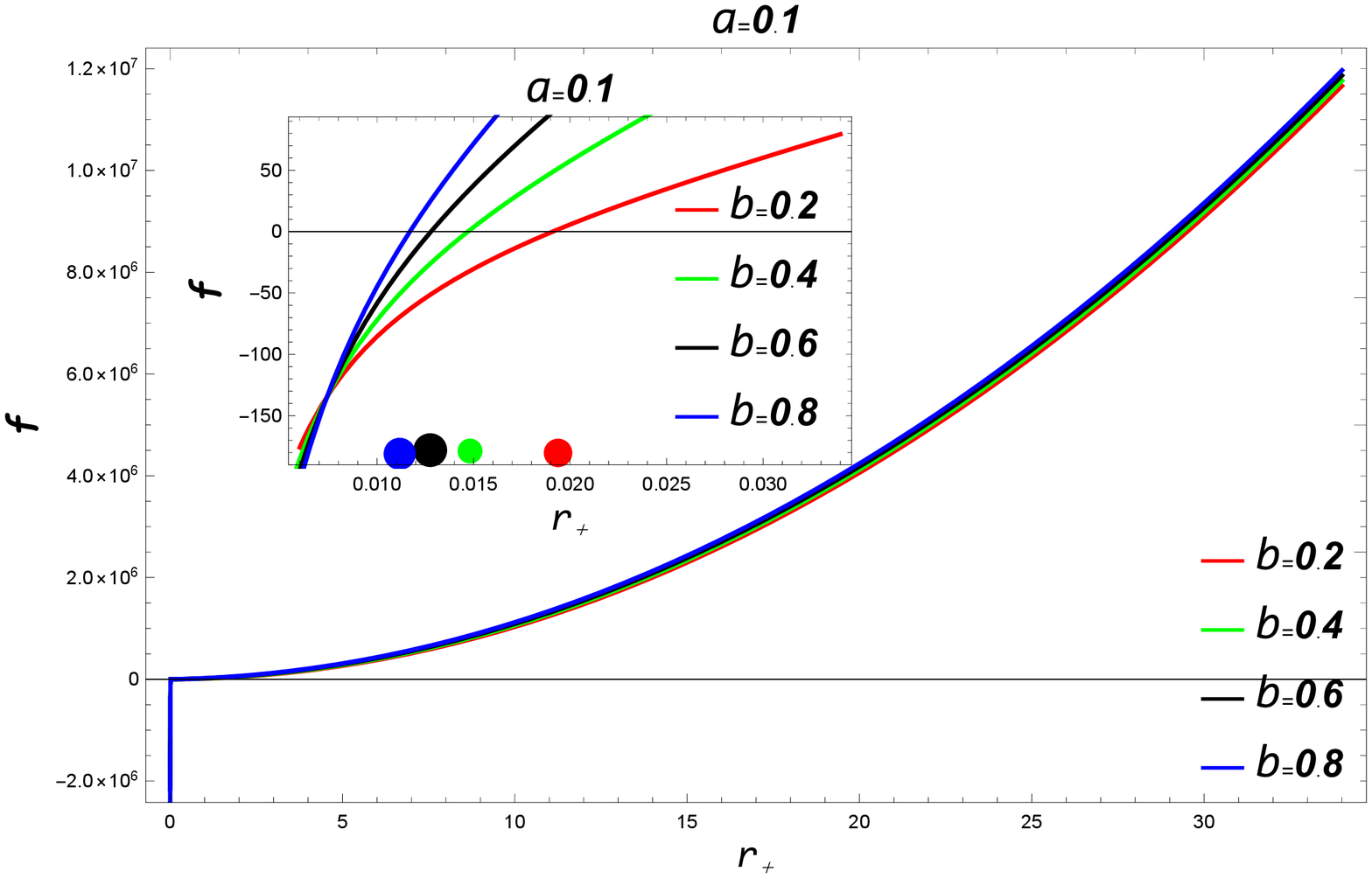}
    \caption{}
     \label{fig:f 2}
      \end{subfigure}
      \hfill
      \begin{subfigure}[b]{.4\textwidth}
    \includegraphics[width=85mm]{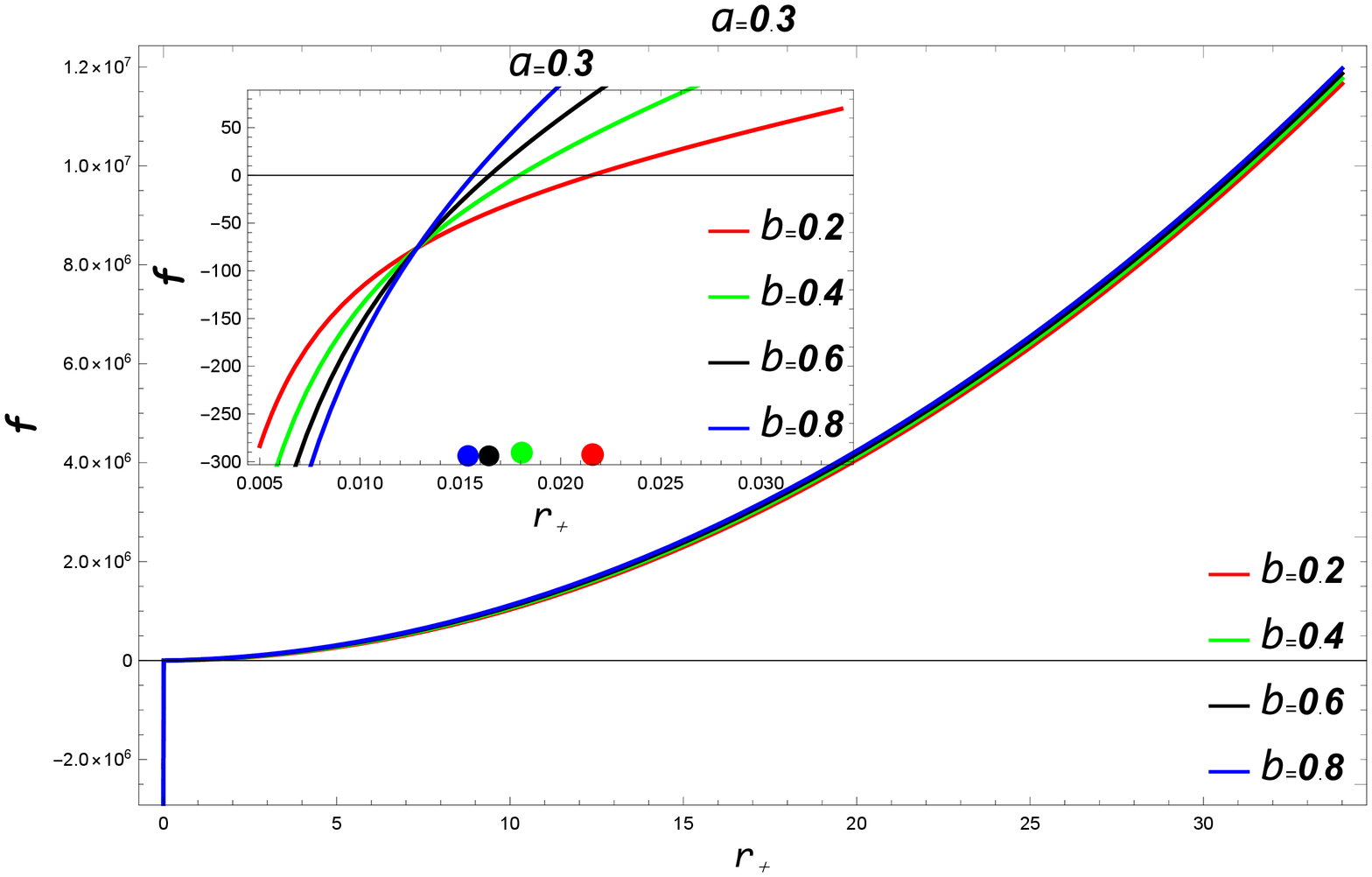}
    \caption{}
     \label{fig:f 2}
      \end{subfigure}
    \vfill
    \centering
    \begin{subfigure}[b]{.4\textwidth}
    \includegraphics[width=85mm]{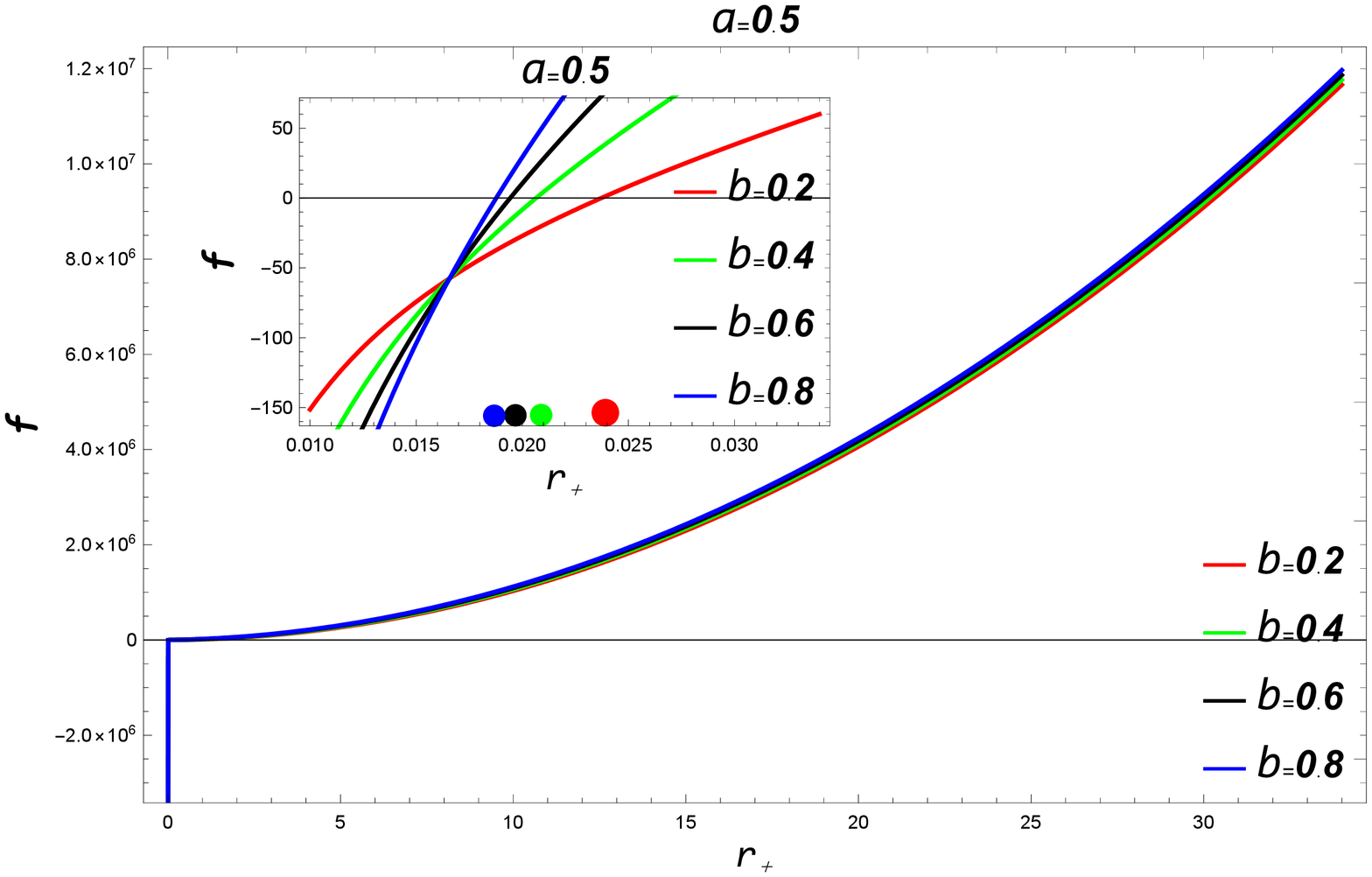}
    \caption{}
     \label{fig:f 2}

   \end{subfigure}
      \hfill
      \begin{subfigure}[b]{.4\textwidth}
    \includegraphics[width=85mm]{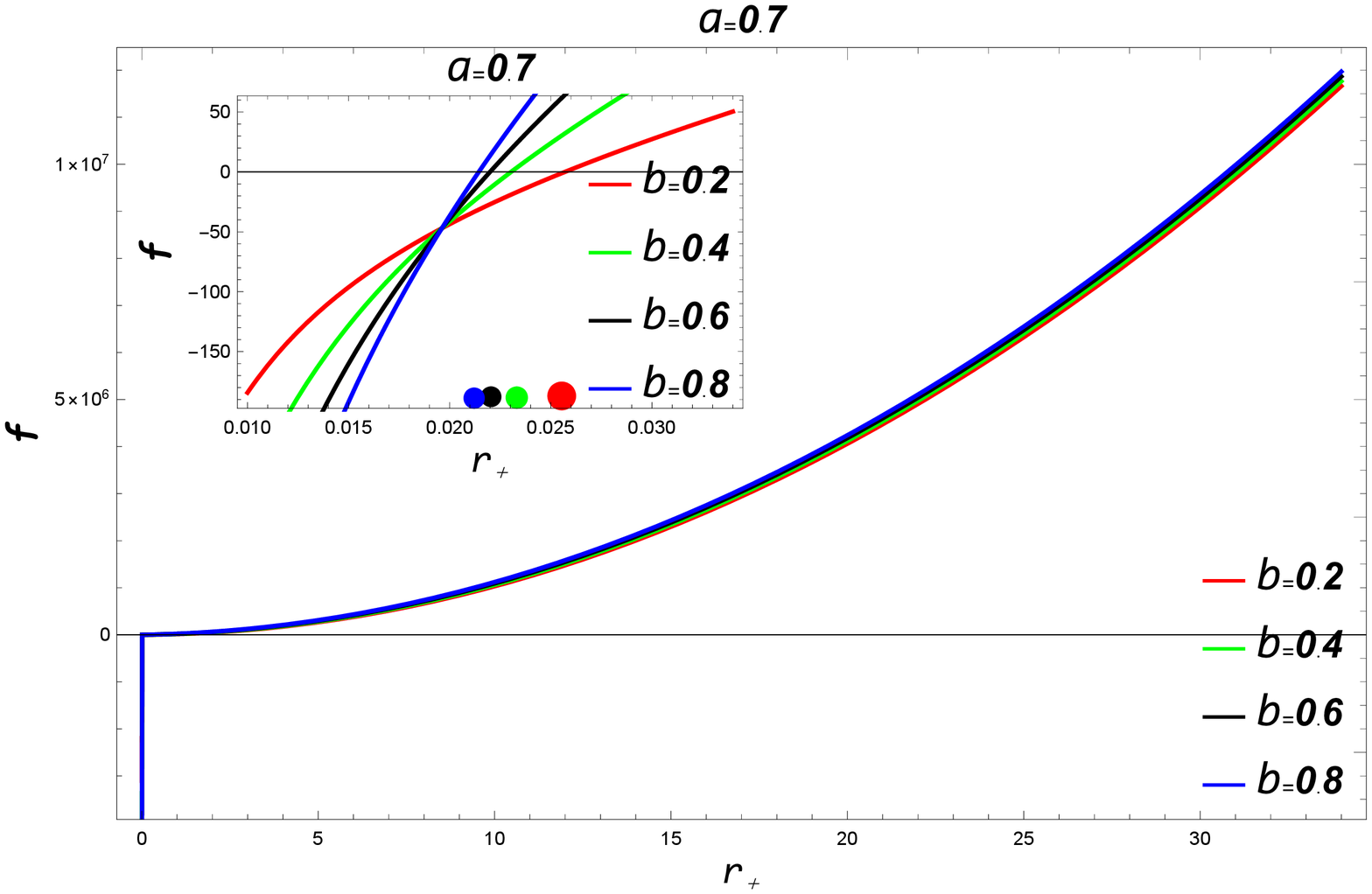}
    \caption{}
     \label{fig:f 2}
      \end{subfigure}

    \caption{The BH horizon, where the horizontal axis denoted by $r_{+}$ and vertical axis by $f$. The pictures for $ a=0.1,a=0.3,a=0.5$  and $a=0.7$  for the different values of $b=0.2, b=0.4,b=0.6,b=0.8.$}
    \label{fig:3}
      \end{figure}

  \begin{longtable}[C]{  c }
 \caption{The  horizon radius for the different values of the parameters $a$ and $b$.\label{long}}\\
\begin{tabular}{ |p{2.5cm}|p{2.5cm}|p{4cm}| }
\hline
   a & b   & $ r_+$ \\
\hline
\multirow{3}{4em}{0.1} & 0.2 & 0.019247\\
& 0.4  &0.014729\\
& 0.6  &0.0128272\\
&0.8 &0.0117176\\
\hline
\multirow{3}{4em}{0.3} & 0.2  &0.02165\\
& 0.4  &0.0179965\\
& 0.6 & 0.0165162\\
& 0.8 & 0.0156372\\
\hline
\multirow{3}{4em}{0.5}&0.2 & 0.023765\\
& 0.4 & 0.020694\\
& 0.6 & 0.019423\\
& 0.8 & 0.0187867\\
\hline
\multirow{3}{4em}{0.7} & 0.2 &0.025927\\
& 0.4 & 0.023051\\

& 0.6 &0.022055\\

& 0.8 & 0.021392\\
\hline
\end{tabular}

\end{longtable}
From the Fig.1 and Table.1, we can see that for the fixed value of
the parameter $a$, the  horizon radius $r_{+}$ decreases by the
increasing value of parameter $b$ but for the fixed value of the
parameter $b$, the horizon $r_{+}$ increases by the increasing
value of parameter $a$.

\section{   Shadow of Van der Waals  black hole}
In this section, we study the shadow of Van der Waals  BH
with the absence of plasma medium.

In order to describe the boundary of the unstable photon circular
orbit, one has to satisfies the effective potential critical
conditions
\begin{equation}\label{15}
V_{eff}=0=\frac{\partial V_{eff}}{\partial r}|_{r_0},
  \frac{\partial^2 V_{eff}}{\partial r^2}|_{r_0}>0
\end{equation}

Using (\ref{14}) and solving the equation (\ref{15}), we obtain
the equation of unstable photon circular orbit in the form
\begin{equation}\label{16}
\frac{ f ^{\prime}(r)}{f(r)}=\frac{2}{r}
\end{equation}

The Photon sphere radius $r_c$ can be obtained graphically, from
the function
\begin{equation}\label{17}
F(r)=r f ^{\prime}-2f(r)=0
\end{equation}
\begin{figure}
 \begin{subfigure}[b]{.4\textwidth}
    \includegraphics[width=85mm]{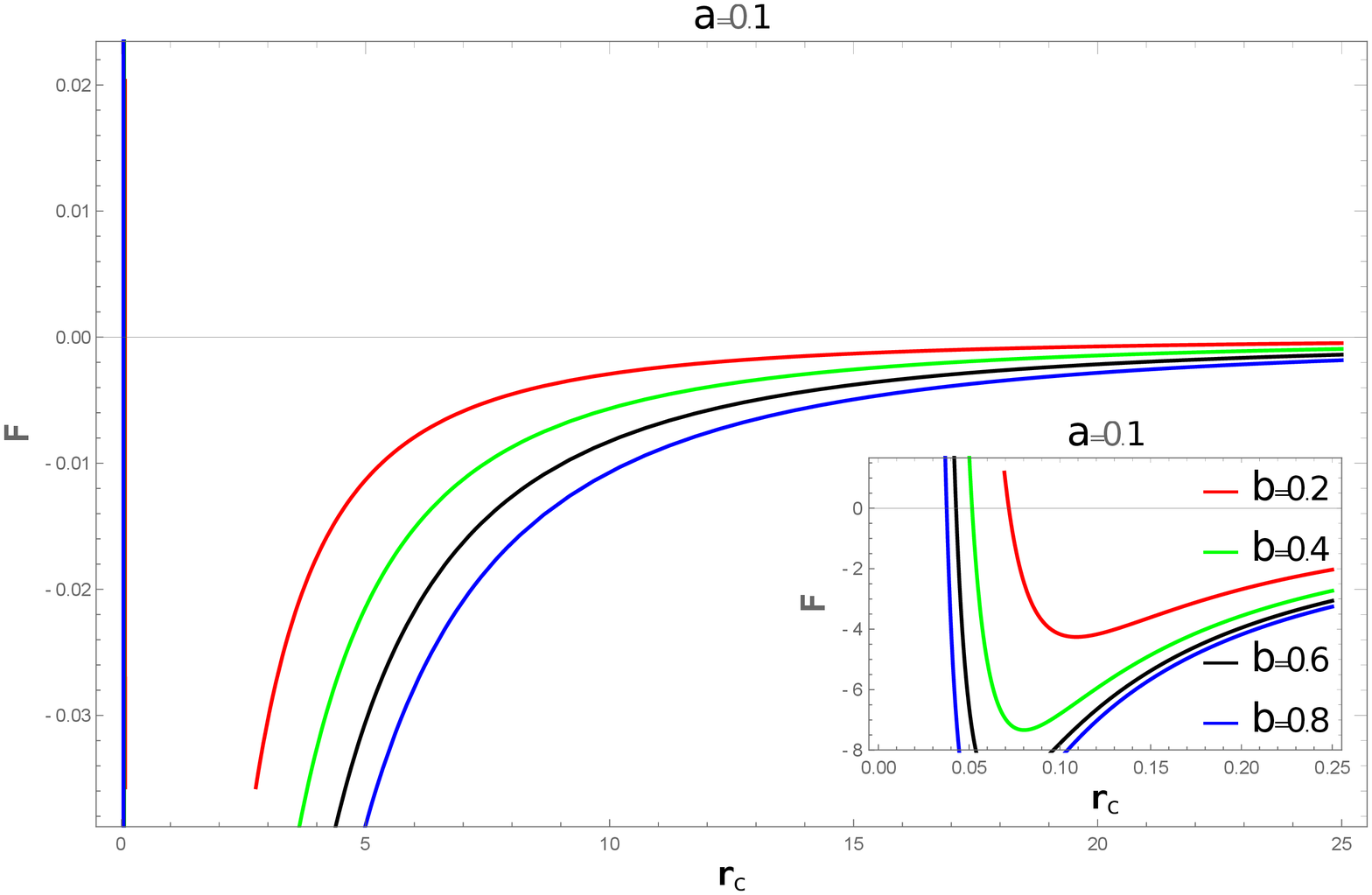}
    \caption{}
     \label{fig:f 2}
      \end{subfigure}
      \hfill
      \begin{subfigure}[b]{.4\textwidth}
    \includegraphics[width=85mm]{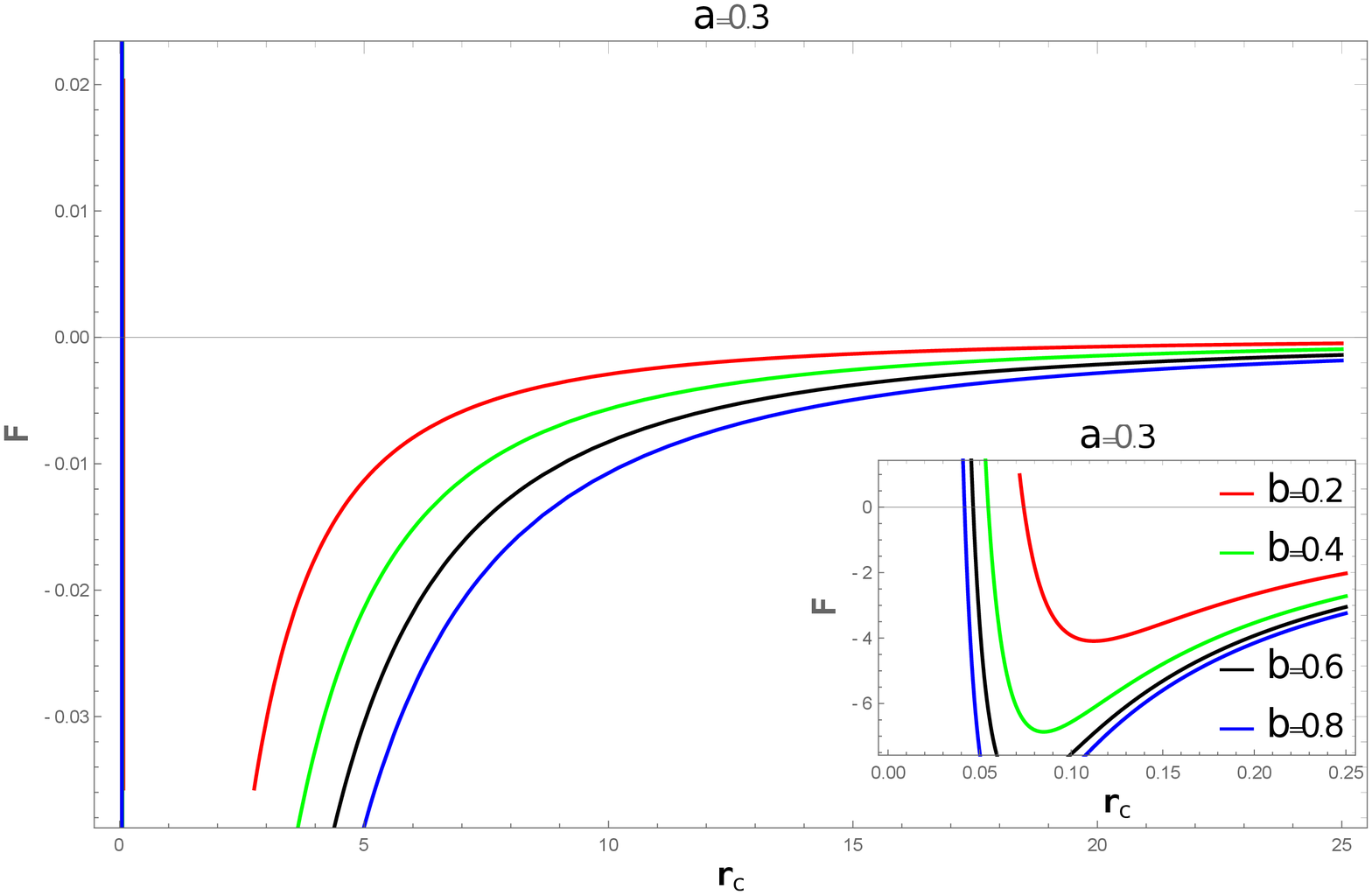}
    \caption{}
     \label{fig:f 2}
      \end{subfigure}
    \vfill
    \centering
    \begin{subfigure}[b]{.4\textwidth}
    \includegraphics[width=85mm]{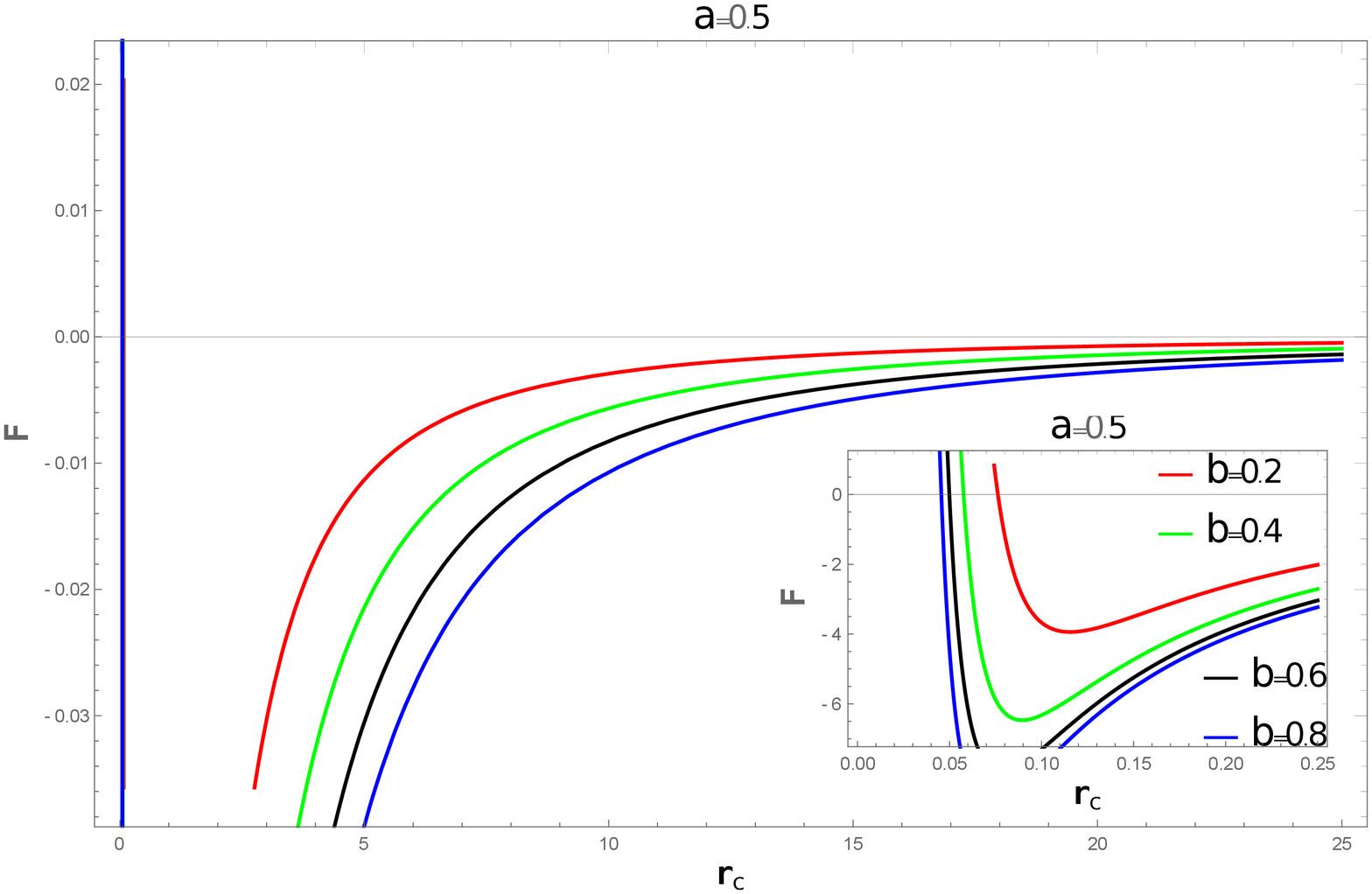}
    \caption{}
     \label{fig:f 2}

   \end{subfigure}
      \hfill
      \begin{subfigure}[b]{.4\textwidth}
    \includegraphics[width=85mm]{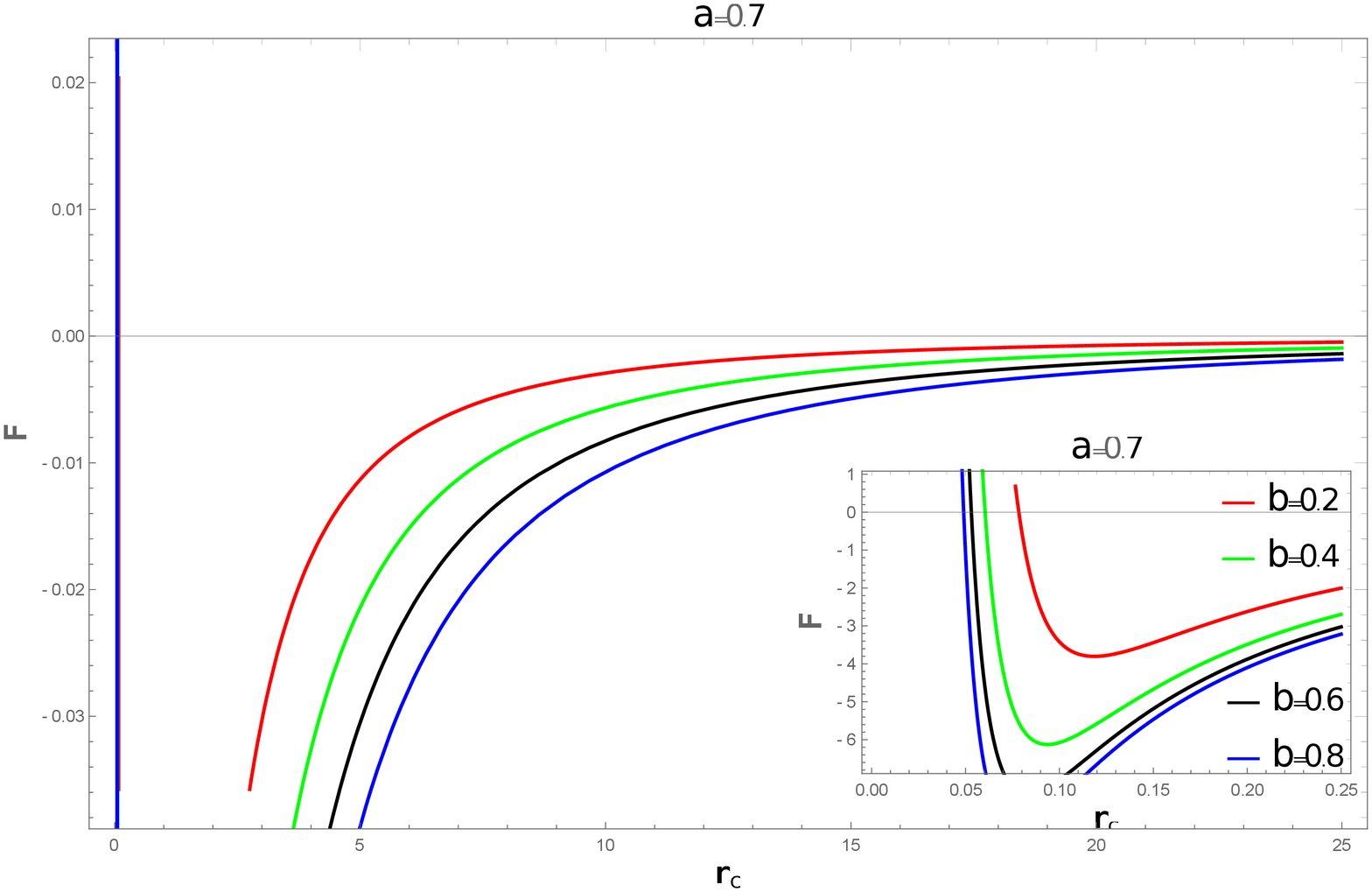}
    \caption{}
     \label{fig:f 2}
      \end{subfigure}
   \caption{The  photon sphere radius $\mathit{r_c}$, where the horizontal axis denoted by $r_c$ and the vertical axis by $F$. The pictures for $ a=0.1,a=0.3,a=0.5$  and $a=0.7$  for the different values of $b=0.2, b=0.4,b=0.6,b=0.8.$}
    \label{fig:3}
      \end{figure}

 \begin{longtable}[C]{  c }
 \caption{Estimation of radius of the photon sphere and the radius of the shadows for the different values of the parameter $a$ and $b$ .\label{long}}\\
\begin{tabular}{ |p{2.5cm}|p{2.5cm}|p{4cm}|p{4cm}| }
\hline
a & b  & $r_{c}$ &$ r_s$ \\
\hline
\multirow{3}{4em}{0.1} & 0.2 & 0.07215 &0.018129\\
& 0.4 & 0.05146 &0.012494\\
& 0.6 & 0.04255 &0.010087\\
&0.8 & 0.03745 &0.008743\\
\hline
\multirow{3}{4em}{0.3} & 0.2 & 0.07419 &0.018315\\
& 0.4 & 0.05460 &0.012746\\
& 0.6 & 0.04639 &0.010347\\
& 0.8 & 0.04137 &0.008921\\
\hline
\multirow{3}{4em}{0.5} & 0.2 & 0.07650 &0.018774\\
& 0.4 & 0.05762 &0.013340\\
& 0.6 & 0.05046 &0.011093\\
& 0.8 & 0.04525 &0.009629\\
\hline
\multirow{3}{4em}{0.7} & 0.2 & 0.07844 &0.019163\\
& 0.4 & 0.06051 &0.013897\\

& 0.6 & 0.05344 &0.011653\\

& 0.8 & 0.04965 &0.010373\\
\hline
\end{tabular}

\end{longtable}

The radius of Shadow $r_s$ at the Photon sphere radius $r_c$  is given by

\begin{equation}\label{18}
r_s=\sqrt{\xi^2 +\eta}=\frac{r_c}{\sqrt{f(r_c)}}
\end{equation}

The shadow boundary curve of the BH at the celestial
co-ordinate ($\alpha,\beta$) define as
\begin{equation}\label{19}
\alpha=\lim_{r_0\rightarrow \infty}(r_0^2  \sin\theta_0)\frac{d\phi}{dr}
\end{equation}

\begin{equation}\label{20}
\beta=\lim_{r_0\rightarrow \infty}(r_0^2  \frac{d\theta}{dr})
\end{equation}
where $r_0$ is the radial distance of the BH with respect
to the observer and $\theta_0$ is the inclination angle between
the BH and observer. With the help of the equations
((\ref{10}),(\ref{11})and (\ref{12}) and doing some
simplification, one can obtain the celestial co-ordinates as

 \begin{equation}\label{21}
\alpha=-\frac{\xi}{\sin\theta_0},
\end{equation}

\begin{equation}\label{22}
\beta=\sqrt{\eta-\frac{\xi^2}{\tan^2\theta_0}}
\end{equation}

Simplifying the equations (\ref{21}) and (\ref{22}),we obtain the following equation
 \begin{equation}\label{23}
 \alpha^2 +\beta^2= \xi^2 +\eta
 \end{equation}

  \begin{figure}
 \begin{subfigure}[b]{.4\textwidth}
    \includegraphics[width=80mm]{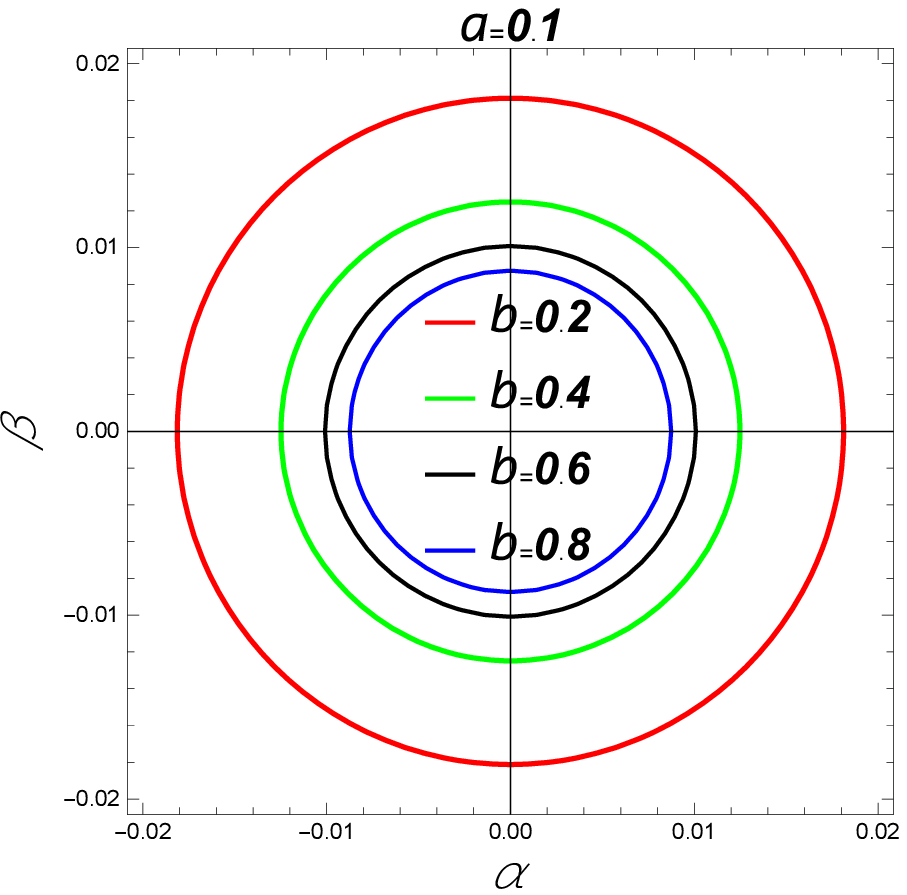}
    \caption{}
     \label{fig:f 2}
      \end{subfigure}
      \hfill
      \begin{subfigure}[b]{.4\textwidth}
    \includegraphics[width=80mm]{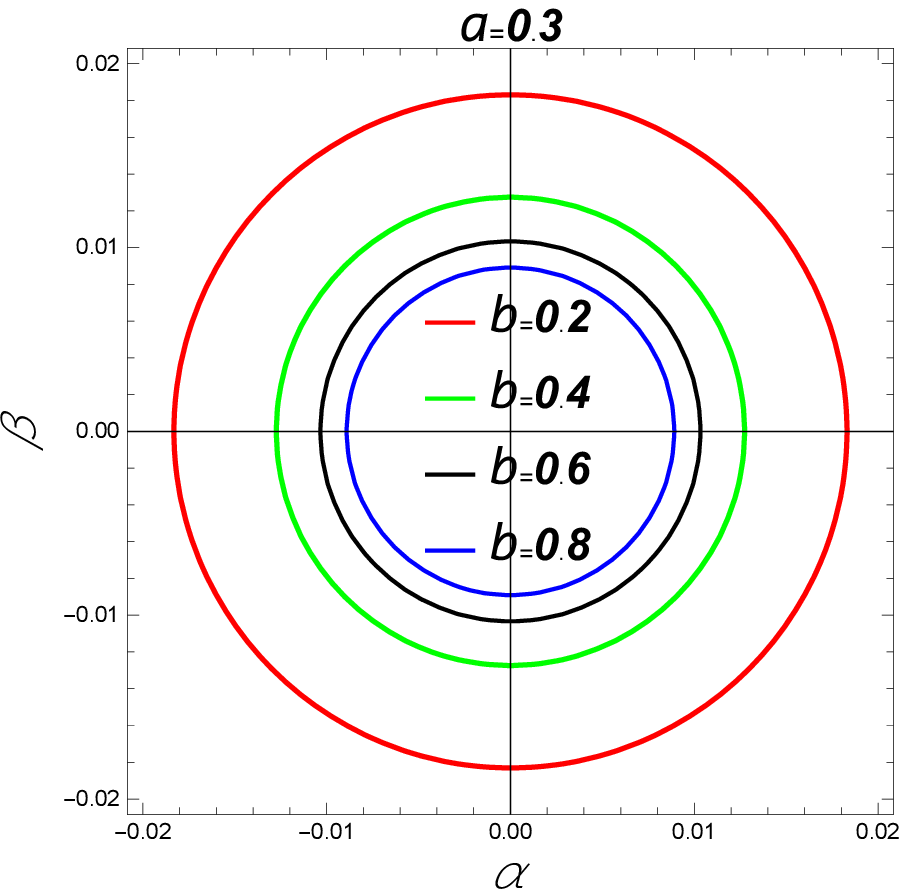}
    \caption{}
     \label{fig:f 2}
      \end{subfigure}
    \vfill
    \centering
    \begin{subfigure}[b]{.4\textwidth}
    \includegraphics[width=90mm]{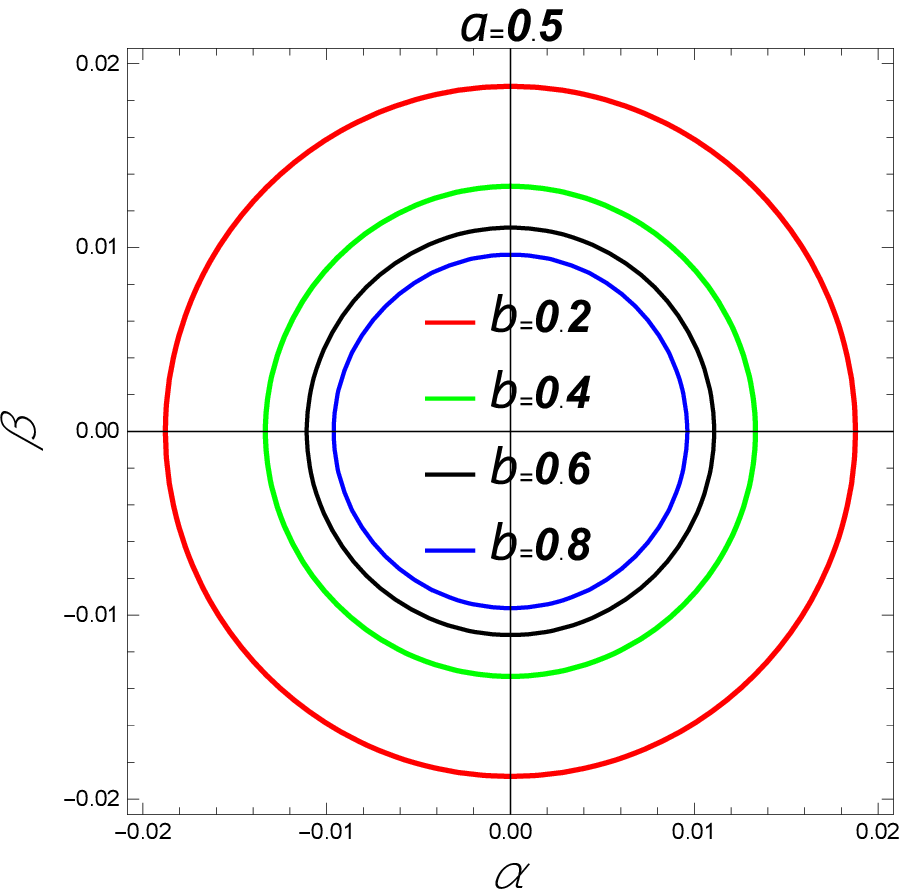}
    \caption{}
     \label{fig:f 2}

   \end{subfigure}
      \hfill
      \begin{subfigure}[b]{.4\textwidth}
    \includegraphics[width=80mm]{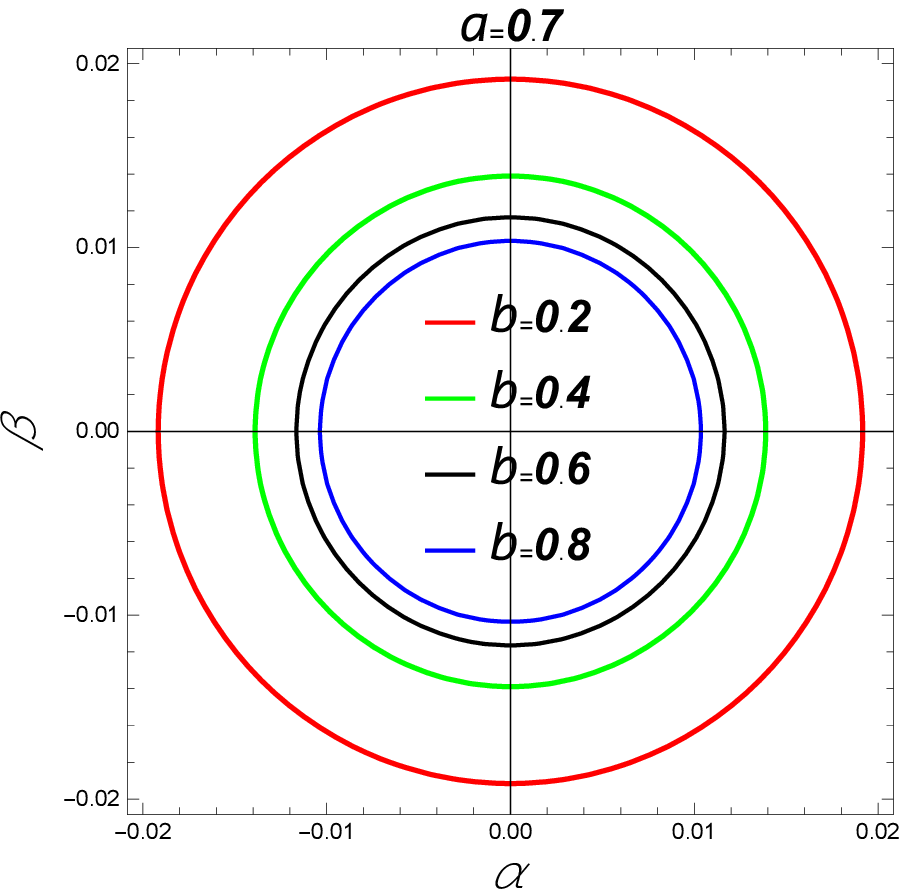}
    \caption{}
     \label{fig:f 2}
      \end{subfigure}

    \caption{The BH shadows in the absence of plasma medium, where the horizontal axis denoted by $r_c$ and the vertical axis by F. The pictures for $ a=0.1,a=0.3,a=0.5$  and $a=0.7$  for the different values of $b=0.2, b=0.4,b=0.6,b=0.8.$}
    \label{fig:3}
      \end{figure}

Using the equations (\ref{18}) and \ref{23}), we construct the
shape of the VdW BH shadow. We plot the shadow shape with $\alpha$
versus $\beta$  for the different values of the parameters $a$ and
$b$ in Fig.3. From Fig.3 and Table.2, we can see that  for the
fixed value of the parameter $a$, the  photon sphere radius $r_c$
and  the radius of the shadow of  BH  $r_s$ is decreased by
increasing the value of parameter b but for the fixed value of the
parameter $b$, the photon sphere radius $r_c$ and  the radius of
the shadow of  BH $r_s$ is increased by increasing the value of
parameter $a$.

 \section{ Effect of homogeneous Plasma on Shadow of Van der Waals black hole}

In this section, we investigate the Van der Waals   BH
shadow in the presence of homogeneous plasma medium. So, We consider
that the spacetime is filled with a homogeneous plasma . If
we denote the mass and the charge of the electron $m$ and $e$
respectively, then the electron plasma frequency is given by
  \begin{equation}\label{24}
  \omega_p(r)^2=\frac{4\pi e^2}{m}N(r)
  \end{equation}
where  electron number density $N(r)$ as a function of
radius coordinates only. The  refraction index
$n$ is related to the photon frequency, reads as
 \begin{equation}\label{25}
 n^2=1-\frac{\omega_p(r)^2}{\omega^2_{\infty}}=1-\sigma^2
 \end{equation}
where  parameter $\sigma =\frac{\omega_p}{\omega_{\infty}}$
represents the ratio of  plasma frequency and  photon frequency.

It is observe that the photon can propagate through the plasma if
$\omega_p<\omega_{\infty}$ and the photon motion is forbidden
while $\omega_p>\omega_{\infty}$. It is clear that the value
$n=1$ is in the vacuum case.

It is useful to define the function  \cite{P.T.B.3.}
 \begin{equation}\label{26}
  h(r)^2=r^2(\frac{1}{f(r)}-\frac{{\omega_p}^2}{\omega^2_{\infty}} )
       \end{equation}

  With the help of the above equation, one  can calculate radius of the photon sphere $r_c$  as real roots of the equation

   \begin{equation}\label{27}
   \frac{d}{dr}(h(r)^2)=0
   \end{equation}
   \begin{equation}\label{28}
H(r)=   \frac{ 2r}{f(r)} -r^2 \frac{ f ^{\prime}(r)}{(f(r))^2}-
  2r\frac{{\omega_p}^2}{\omega^2_{\infty}}=0
   \end{equation}

The angular radius of the Shadow is defined as follows

   \begin{equation}\label{29}
   \sin^2(\alpha_s)=\frac{r_c^2(\frac{1}{f(r_s)}-\frac{{\omega_p}^2}{\omega^2_{\infty}} )}{r_o^2(\frac{1}{f(r_o)}-\frac{{\omega_p}^2}{\omega^2_{\infty}} )}
   \end{equation}

   For the Vacuum case,$\omega_p=0$

   \begin{equation}\label{30}
   \sin^2(\alpha_s)=\frac{\frac{r_c^2}{f(r_s)}}{\frac{r_o^2}{f(r_o)}}
   \end{equation}

   where   $r_c$ is given from the equation (\ref{26}) graphically.

Solving the equation (\ref{28}), we  graphically obtain the radius
of the photon sphere in the presence of plasma medium and using
this radius, we construct the shape of  shadows of the BH in a
homogeneous plasma medium. We plot the shadow shape with $\alpha$
versus $\beta$ in Fig.5. We can see from Fig.5 and Table.3 that
the radius of the shadow of the BH decreases by increasing the
value of parameter $\sigma$, and we also noticed that the radius
of  the BH shadow in the presence of plasma is bigger than the
vacuum one.

 \begin{figure}
 \begin{subfigure}[b]{.4\textwidth}
    \includegraphics[width=90mm]{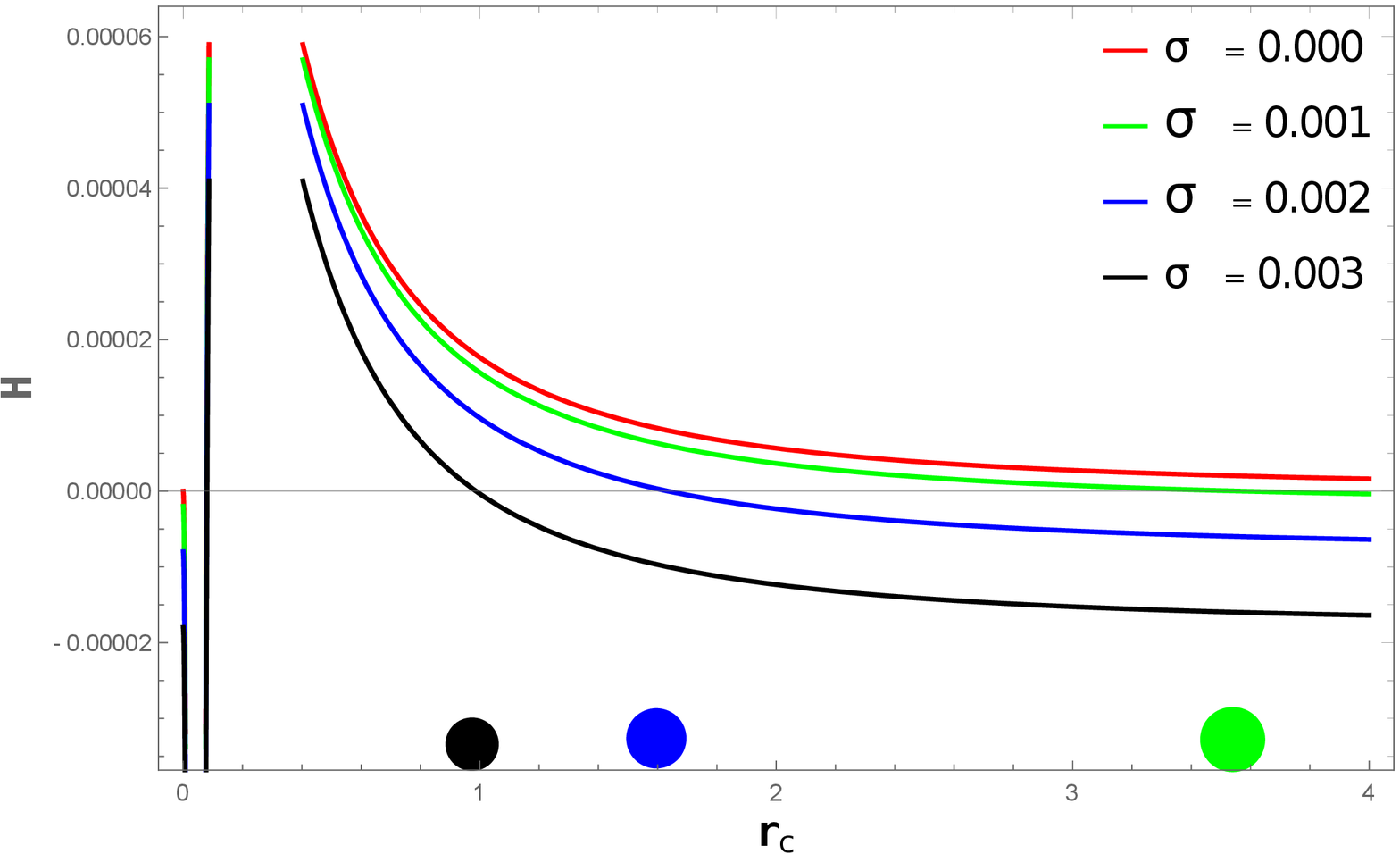}
    \caption{}
     \label{fig:f 2}
      \end{subfigure}
      \hfill
      \begin{subfigure}[b]{.4\textwidth}
    \includegraphics[width=90mm]{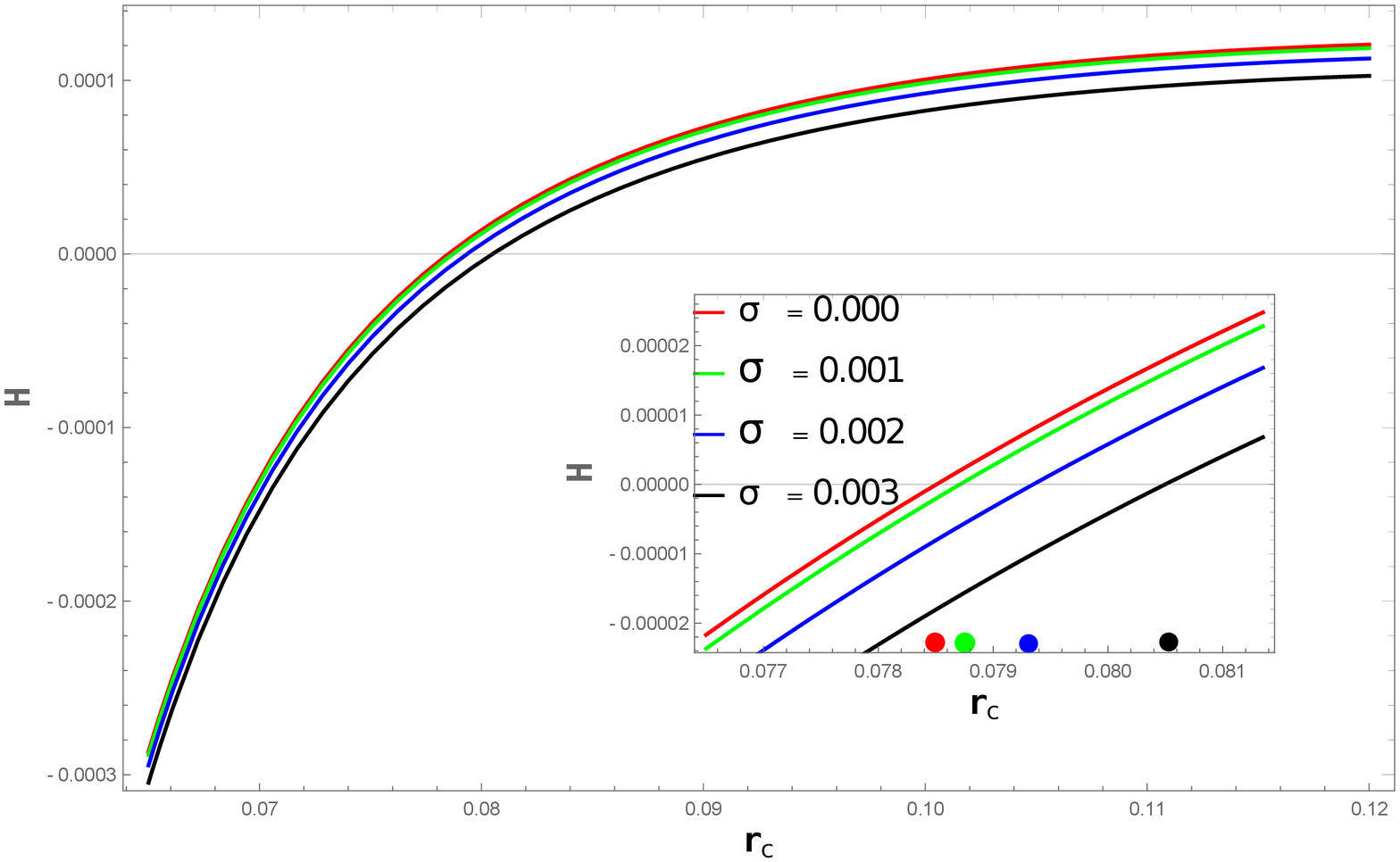}
    \caption{}
     \label{fig:f 2}
      \end{subfigure}
    \vfill
    \centering
    \begin{subfigure}[b]{.4\textwidth}
    \caption{}
     \label{fig:f 2}
      \end{subfigure}
    \caption{The photon sphere radius $\mathit{r_c}$, where the horizontal axis denoted by $r_c$ and the vertical axis by H. Right panel for the case  photon  sphere radius $r_c$ within the range  $ 0<r_c<1$ and Left panel for the case  photon  sphere radius $r_c\geq 1$  .}
    \label{fig:}
      \end{figure}
\begin{figure}
 \begin{subfigure}[b]{.4\textwidth}
    \includegraphics[width=80mm]{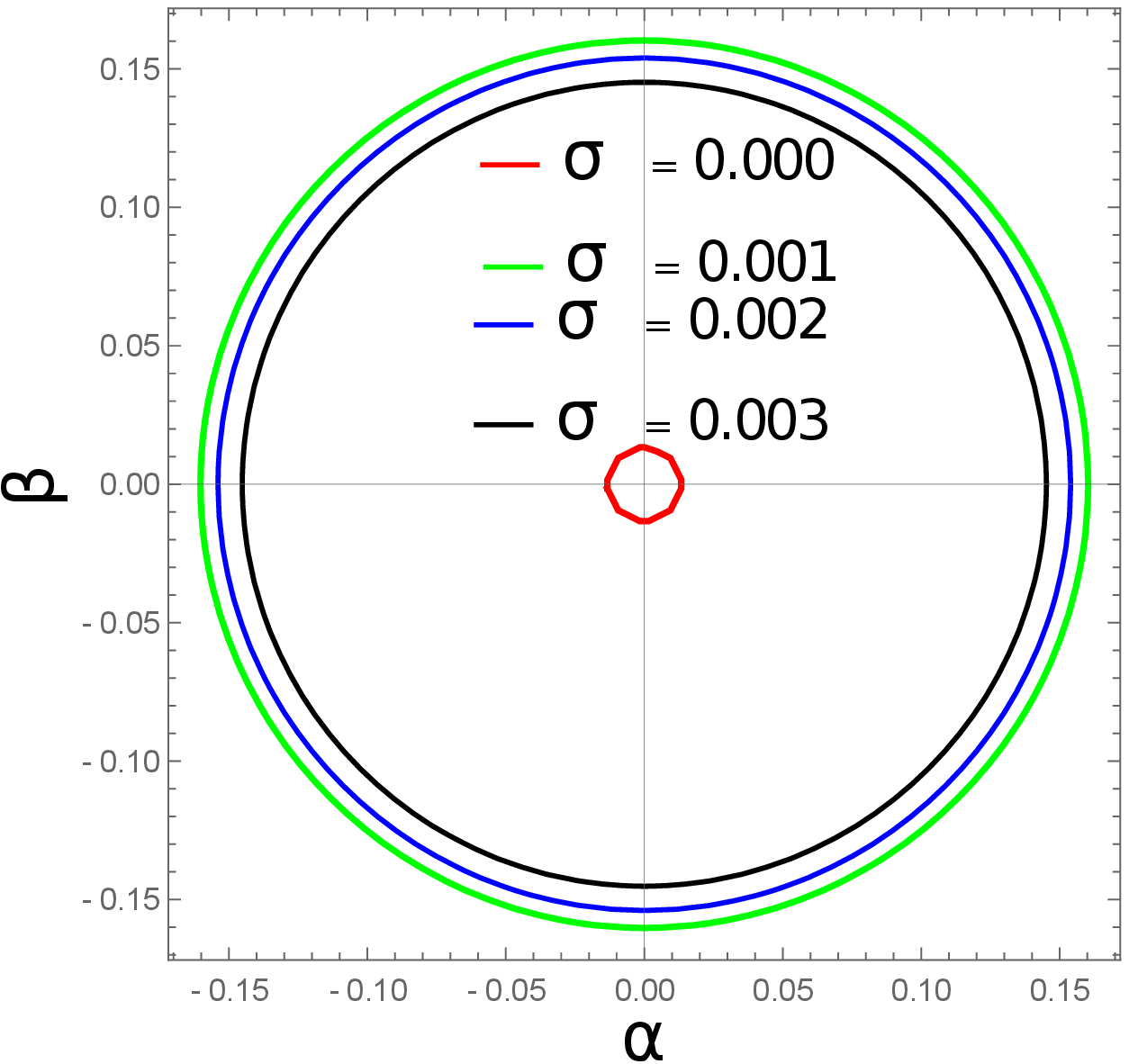}
    \caption{}
     \label{fig:f 2}
      \end{subfigure}
      \hfill
      \begin{subfigure}[b]{.4\textwidth}
    \includegraphics[width=80mm]{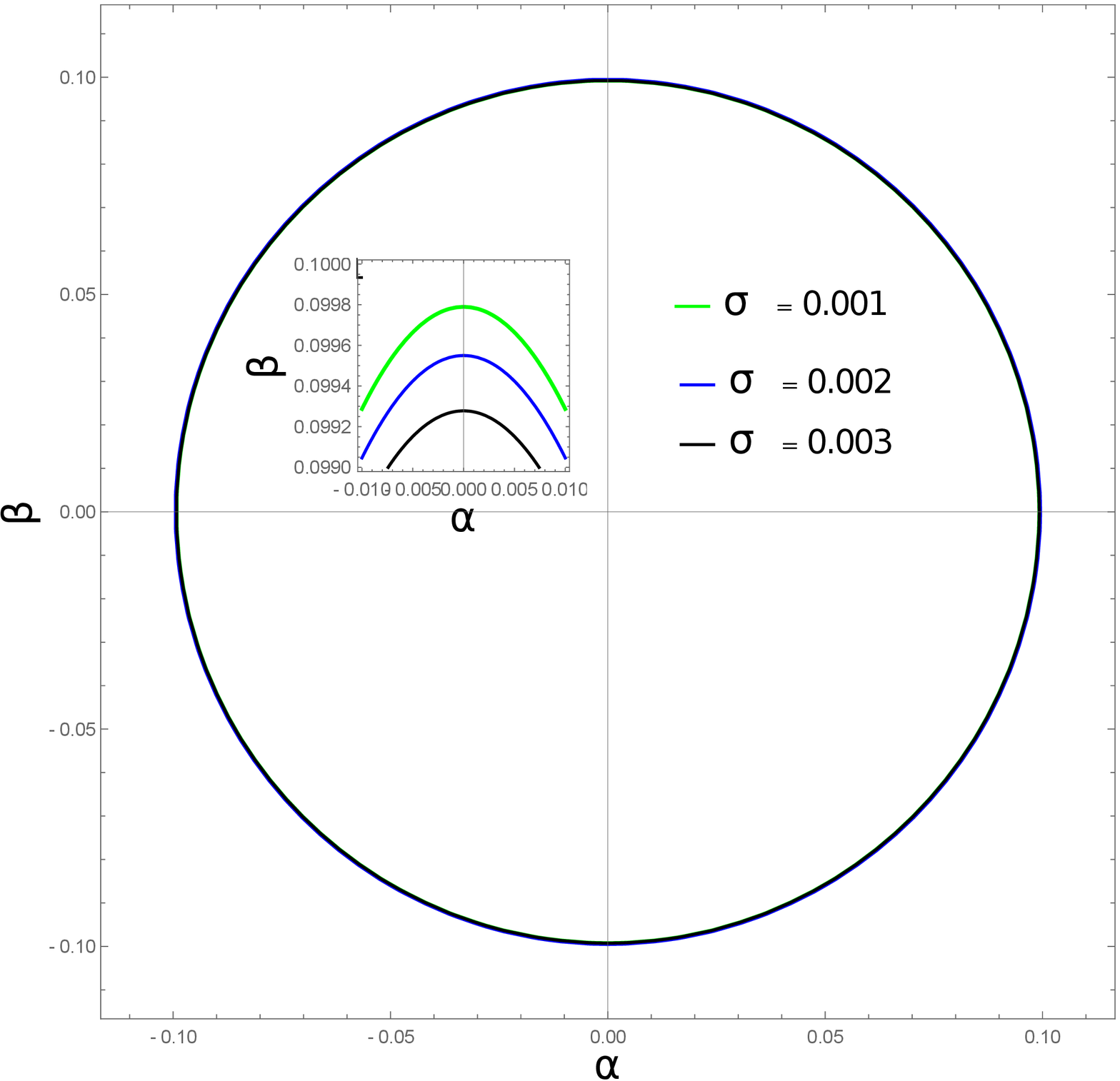}
    \caption{}
     \label{fig:f 2}
      \end{subfigure}
    \vfill
    \centering
    \begin{subfigure}[b]{.4\textwidth}
    \caption{}
     \label{fig:f 2}
      \end{subfigure}
    \caption{The black hole shadows  in the presence of  different plasma medium ($\sigma=0.000,0.001,0.002,0.003 $) , where the horizontal axis denoted by $\alpha$ and the vertical axis by $\beta$.  Parameter $\sigma =\frac{\omega_p}{\omega_{\infty}}$ represents the ratio of plasma frequency to the photon frequency. Left panel for the case  photon  sphere radius $r_c$ within the range  $ 0<r_c<1$ and Right panel for the case  photon  sphere radius $r_c> 1$. In the Left panel , the red circle represents the shadow shape in vacuum medium. }
    \label{fig:}

      \end{figure}
       \begin{longtable}[C]{  c }
 \caption{Estimation  of the photon sphere radius $r_c$ and the shadow radius  of the black hole  $r_s$ for the different values of the parameter $\sigma$.\label{long}}\\
\begin{tabular}{ |p{1.5cm}|p{2cm}|p{2cm}|p{2cm}|  }
\hline
$\sigma= \frac{\sigma_p}{\sigma_{\infty}}$ &$ 0<   r_{c}<1$ & $r_s$\\
\hline
 0.000& 0.078516 & 0.0139318 \\

\hline
0.001&0.078704 & 0.0160289 \\
\hline
0.002& 0.079342 & 0.0153975 \\
\hline
0.003& 0.080528 & 0.0145246 \\
\hline
\end{tabular}
\vline

\begin{tabular}{ |p{1.5cm}|p{2cm}|p{2cm}|p{2cm}|  }
\hline
$\sigma= \frac{\sigma_p}{\sigma_{\infty}}$ &$1<  r_{c}<\infty$ & $r_s$\\
\hline
0.001 & 3.54159 & 0.09978958 \\

\hline
0.002& 1.63751 &0.09954998\\
\hline
0.003& 1.00143 & 0.09927789\\

\hline
\end{tabular}

\end{longtable}
\begin{figure}
 \begin{subfigure}[b]{.4\textwidth}
    \includegraphics[width=80mm]{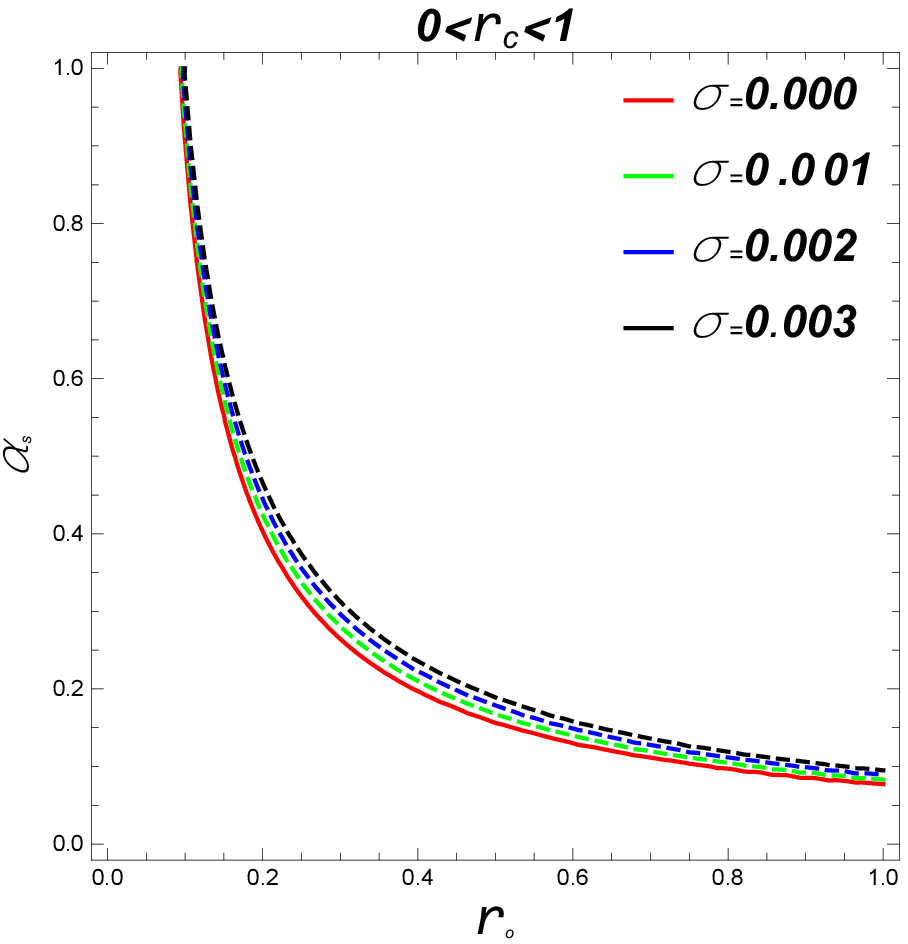}
    \caption{}
     \label{fig:f 2}
      \end{subfigure}
      \hfill
      \begin{subfigure}[b]{.4\textwidth}
    \includegraphics[width=80mm]{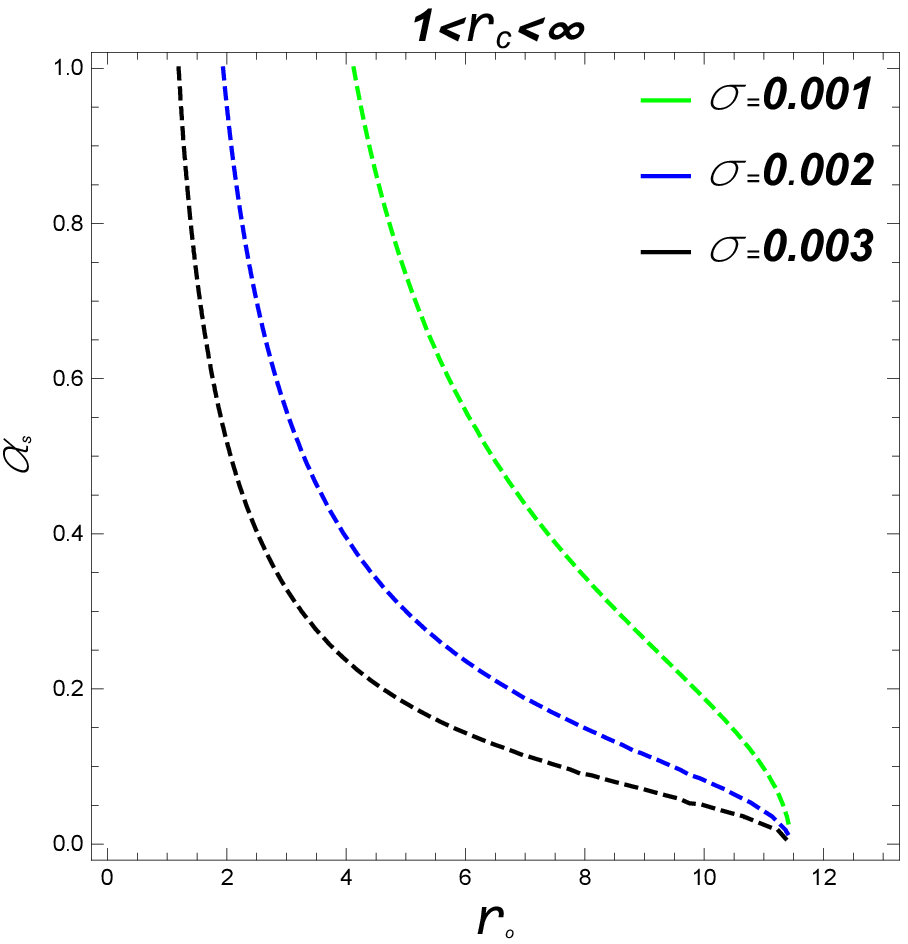}
    \caption{}
     \label{fig:f 2}

      \end{subfigure}
    \caption{Behaviour of angular size of the shadow of BH with respect to the position of  a static observer $r_{o}$ with different value of  $\sigma=0.000,0.001,0.002,0.003$.Left panel represents the angular size of shadow  when  the  photon sphere radius $r_c<1$ and right panel represents the angular size of shadow  when the photon sphere radius $r_c>$.  Parameter $\sigma =\frac{\omega_p}{\omega_{\infty}}$ represents the ratio of  plasma frequency and  photon frequency.}
    \label{fig:}
      \end{figure}
Using the equation (\ref{29})and Table.3, we obtain the angular
size of the shadow of the BH and the variation of its
displayed in Fig.6. In this figure, we see that the angular size
of shadow is decreased with  the position of a static observer
$r_{o}$ for the different value of the parameter $\sigma
=\frac{\omega_p}{\omega_{\infty}}$.

\section{Strong gravitational lensing with the effects of homogeneous plasma}

In this section we discuss the strong GL on the
equatorial plane( $\theta=\frac{\pi}{2}$) in the presence of
homogeneous plasma. The metric (\ref{5}) can be written on the
equatorial plane ($\theta=\frac{\pi}{2}$) as
 \begin{equation}\label{31}
ds^2=-F(r)dt^2+ G(r) dr^2 +K(r) d\phi^2
\end{equation}
where
\begin{equation}\label{32}
  F(r) = f(r)=2\pi a-\frac{2M}{r}+  \frac{r^2}{l^2}(1+\frac{3b}{2r}) -\frac{3\pi a b^2}{r(2r+3b)} -\frac{4\pi a b}{r}log(\frac{r}{b} +\frac{3}{2})
    \end{equation}
\begin{equation}\label{33}
G(r)= ( f(r)^{-1}=\{2\pi a-\frac{2M}{r}+  \frac{r^2}{l^2}(1+\frac{3b}{2r}) -\frac{3\pi a b^2}{r(2r+3b)} -\frac{4\pi a b}{r}log(\frac{r}{b} +\frac{3}{2})\}^{-1}
 \end{equation}

  and
  \begin{equation}\label{34}
  K(r)=r^2
   \end{equation}
The Hamiltonian for the light rays around the Van der Waals BH when it's surrounded by plasma can be expressed as in Ref.\cite{K.L.}

   \begin{equation}\label{35}
   H(x^i,p_i)=\frac{1}{2}[g^{ik}P_iP_k+ w^2_{p}]
   \end{equation}

where $ g^{ik}$is the contravariant tensor of the metric
(\ref{31}) and $P_i$ is the momentum of photon.

Here, we consider the  homogeneous  plasma with
$\omega_p=constant$. Using the equation\ref{35}) , we can obtain the Hamiltonian differential equations for the photon around the black hole as

   \begin{equation}\label{36}
    \frac{dx^{i}}{d\lambda}=\frac{\partial{H}}{\partial{p_i}},
   \frac{dx^{i}}{d\lambda}=-\frac{\partial{H}}{\partial{x^i}}
   \end{equation}

  and hence, we obtain two constants of motions which are the angular momentum $L$ and the energy $E$ of the photon

   \begin{equation}\label{37}
   L=P_{\phi}, E=-p_t=\omega_{\infty},
   \end{equation}

   where $\omega_{\infty}$ is the  frequency  of photon coming from infinity.

 Using the Eqs. (\ref{31}) and (\ref{36}), the expression for $dr/d\lambda, d\phi/d\lambda$ can written as

  \begin{equation}\label{38}
   \frac{dr}{d\lambda}=\frac{\partial{H}}{\partial{p_i}}=\frac{p_{r}}{G(r)}
    \end{equation}
    \begin{equation}\label{39}
   \frac{d\phi}{d\lambda}=\frac{\partial{H}}{\partial{x^i}}=\frac{p_{\phi}}{K(r)}
   \end{equation}

Using the Eqs (\ref{38}) and (\ref{39}),one can find  the equation of  photon  trajectory  as in Ref. \cite{J.G.L,T.16.}
\begin{equation}\label{40}
(\frac{dr}{d\phi})^2=\frac{R_p(r)K(r)}{G(r)}
\end{equation}

 Where
 \begin{equation}\label{41}
 R_p=\frac{E^2K(r)W(r)}{L^2 F(r)}-1
 \end{equation}

 and
 \begin{equation}\label{42}
 W(r)=1-\frac{\omega^2_p(r) F(r)}{E^2}=1-\frac{\omega^2_p(r) }{\omega_{\infty
 }^2}F(r)=1-\sigma^2F(r)
 \end{equation}
where parameter $\sigma =\frac{\omega_p}{\omega_{\infty}}$
represents the ratio of  plasma frequency and  photon frequency.

In particular, $\omega_p(r) = 0 $ as well as $ W(r) = 1,$ in  the
Eq. (\ref{42})  represent the motion of photon ray in vacuum
medium.

As the photon rays coming from infinity and reaches to  closest
distance $ r = r_0$ and then goes out to infinity. So, for the
minimum distance of photon trajectory, $\frac{dr}{d\phi} $
vanishes and hence, we have the minimum impact parameter  as
\cite{T.4.}
\begin{equation}\label{43}
u(r_0)= \frac{L}{E}=\frac{K(r_0)}{F(r_0)} \frac{\dot{\phi}}{\dot{t}}=\sqrt{\frac{K(r_0)W(r_0)}{F(r_0)}}
\end{equation}
To find the photon sphere radius for the unstable circular
orbit of photon for static,spherically symmetric BH, we consider a function $h(r)$ given by ref. \cite{P.T.B.3.} as
\begin{equation}\label{44}
(h(r))^2=\frac{W(r)K(r)}{F(r)}=\frac{K(r)}{F(r)}\{1-\frac{`\omega^2_p F(r)}{E^2}\}
\end{equation}

The  photon sphere radius $r_c$ is the largest real root of
the following equation
\begin{equation}\label{45}
\frac{d}{dr}\{(h(r))^2\}=0
\end{equation}

  \begin{figure}
 \begin{subfigure}[b]{.4\textwidth}
    \includegraphics[width=80mm]{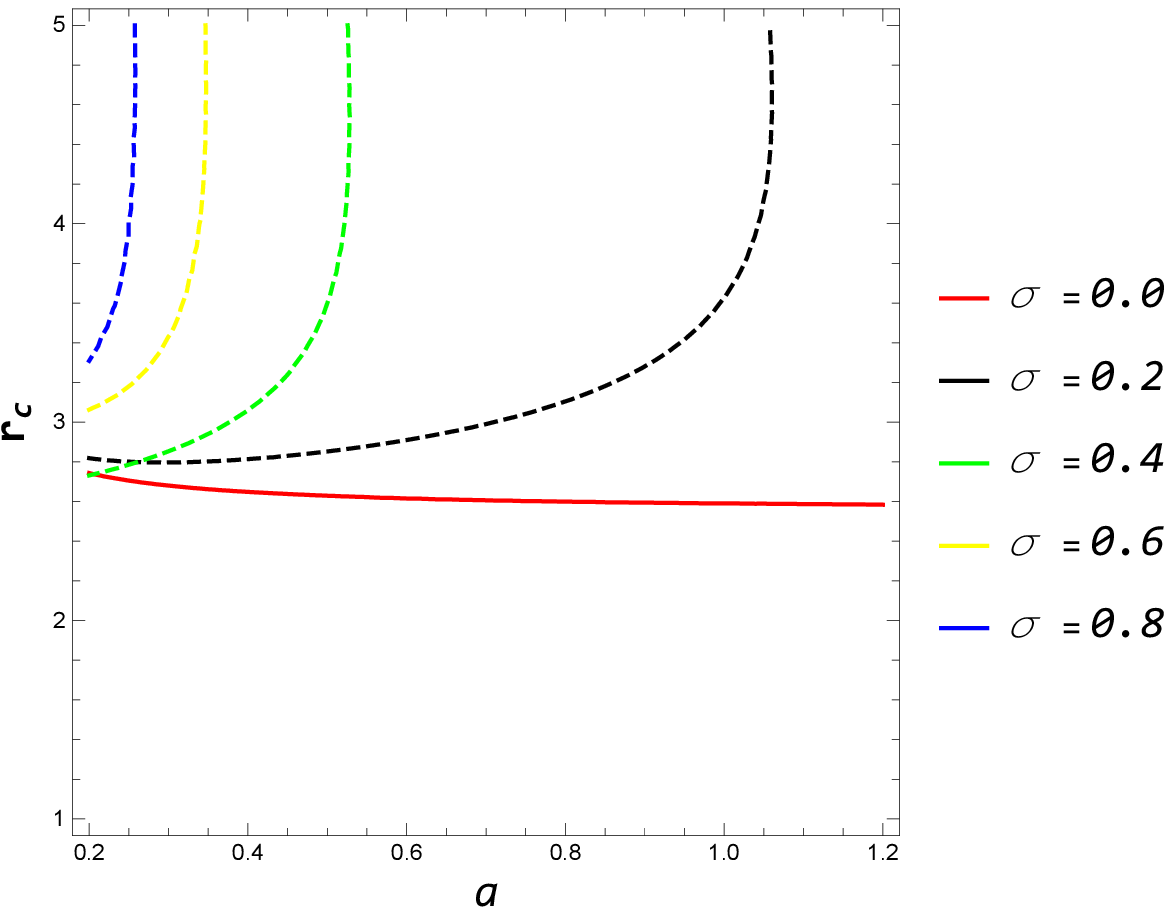}
    \caption{}
     \label{fig:f 2}
      \end{subfigure}
      \hfill
      \begin{subfigure}[b]{.4\textwidth}
    \includegraphics[width=80mm]{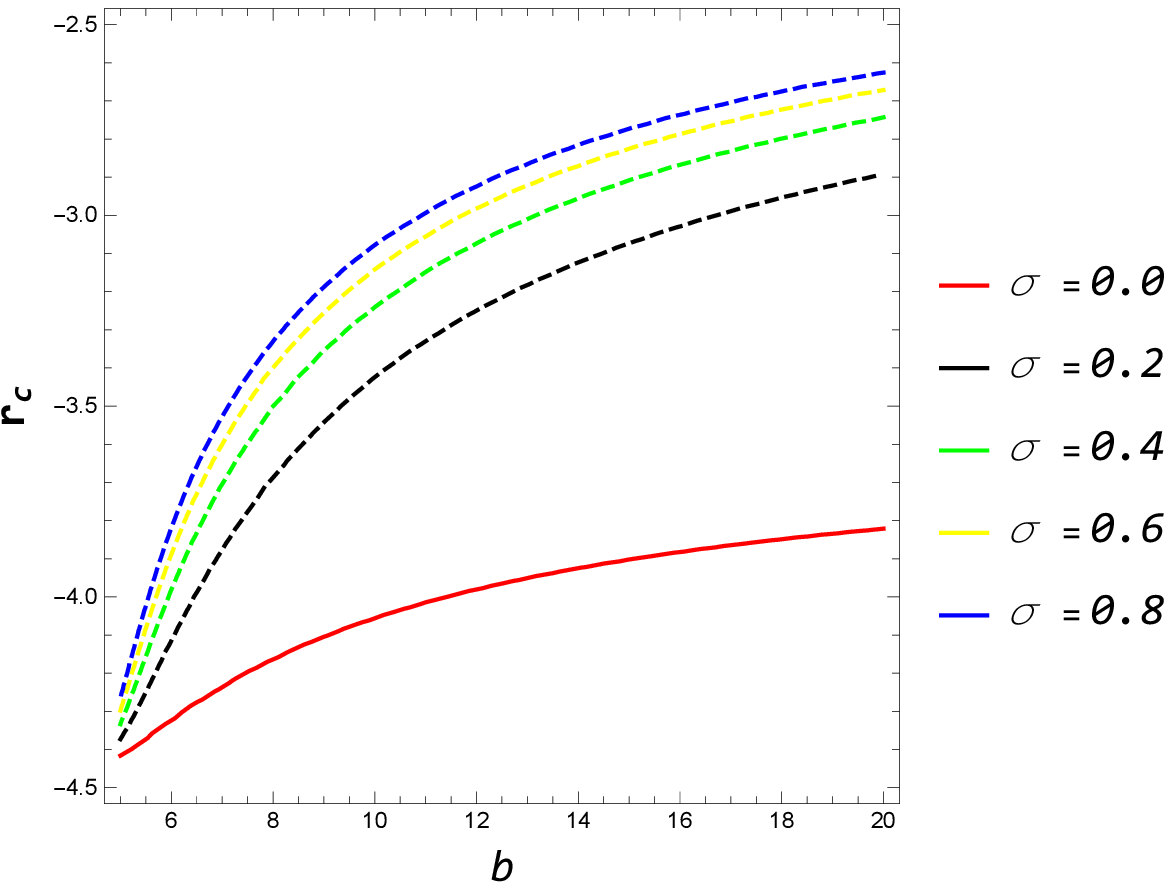}
    \caption{}
     \label{fig:f 2}

      \end{subfigure}
    \caption{The radius $\mathit{r_c}$ as a function of  $a$ (Left panel) and  $b$ (Right panel)for the  values of $\sigma=0.0,0.2,0.4,0.4,0.6,0.8$. Parameter $\sigma =\frac{\omega_p}{\omega_{\infty}}$ represents the ratio of plasma frequency to the photon frequency.}
    \label{fig:}
      \end{figure}

\begin{figure}
 \begin{subfigure}[b]{.4\textwidth}
    \includegraphics[width=80mm]{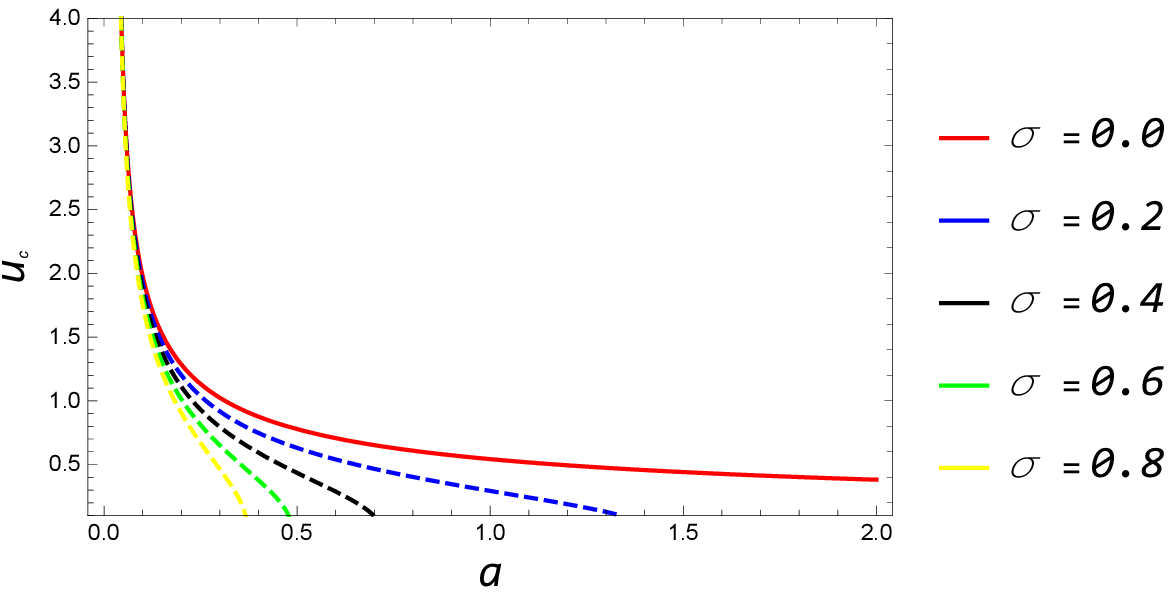}
    \caption{}
     \label{fig:f 2}
      \end{subfigure}
      \hfill
      \begin{subfigure}[b]{.4\textwidth}
    \includegraphics[width=80mm]{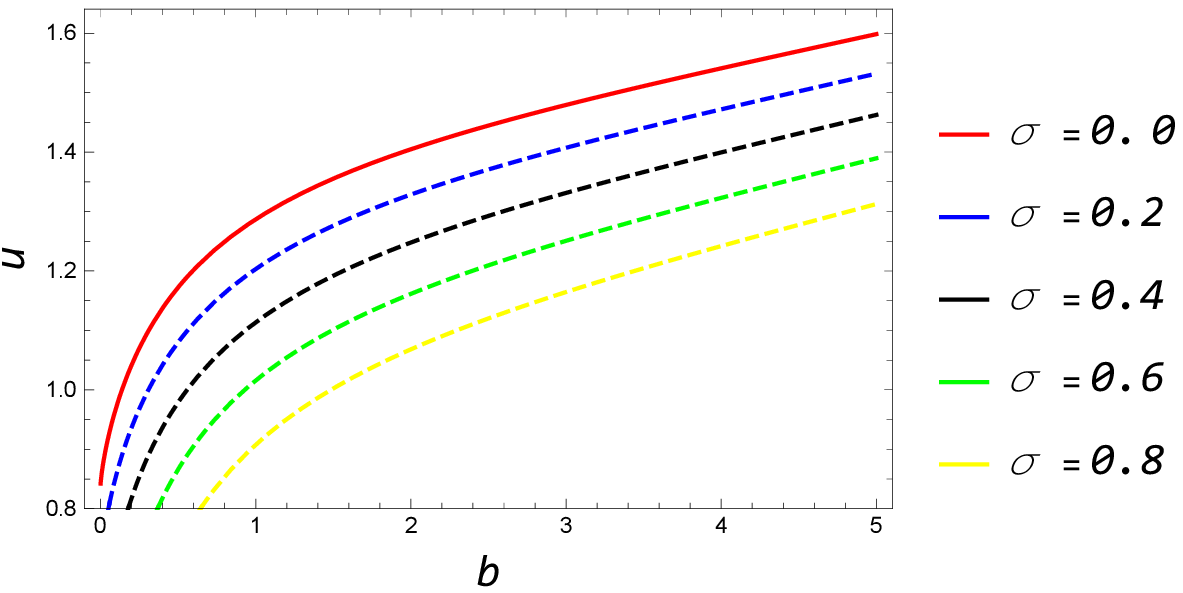}
    \caption{}
     \label{fig:f 2}

      \end{subfigure}
    \caption{The  impact parameter $\mathit{u_c}$ as a function of  $a$  (Left panel) and  $b$  (Right panel)for the  values of $\sigma=0.0,0.2,0.4,0.4,0.6,0.8$. Parameter $\sigma =\frac{\omega_p}{\omega_{\infty}}$ represents the ratio of  plasma frequency and  photon frequency.}
    \label{fig:}
      \end{figure}

Simplifying the equations (\ref{44}) and (\ref{45}), we obtain,

\begin{equation}\label{46}
\frac{K^{\prime}}{K}+\frac{W^{\prime}}{W}-\frac{F^{\prime}}{F}=0
\end{equation}
where $\prime$ denotes the differentiation with respect to $r$.

One can solve the equation (\ref{46}), we  obtain the photon
sphere radius $r_c$ and it's behaviour is shown in Fig.7.The
photon sphere radius $r_c$ increases with the parameter $a$ (left
panel) as well as the parameter $b$ (right panel) in Fig.7. We can
see that The photon sphere radius $r_c$ in a homogeneous plasma
medium is greater than  the vacuum one.

We define minimum impact parameter $u_c$ at $ r_0 = r_c$, for the light ray as

\begin{equation}\label{47}
u(r_c)=\sqrt{\frac{K(r_c)W(r_c)}{F(r_c)}}
\end{equation}
which is shown in Fig.8. In this figure, we can see that  for the
fixed value of the parameter $\sigma$, the  minimum impact
parameter $u_c$ decreases with the parameter $a$ (left panel) and
increases the parameter $b$ (right panel). We also can see that
minimum impact parameter $u_c$ in a homogeneous plasma medium is
smaller than the vacuum case.

From the Eq.(\ref{40}), the angle of deflection $\alpha(r_0)$ for the light ray coming from infinity to the Van der Waals BH in
a homogeneous plasma medium can be written as

\begin{equation}\label{48}
\alpha(r_0)=\phi(r_0)- \pi
\end{equation}
where
\begin{equation}\label{49}
\phi(r_0)=2 \int _{r_0}^\infty  \frac{d\phi}{dr} dr
\end{equation}
\begin{equation}\label{50}
 \phi(r_0)= 2 \int_{r_0}^\infty \sqrt{\frac{ G(r)}{R_p(r)K(r) }}dr
 \end{equation}

To calculate the   angle of deflection in strong field limit, we apply an useful method developed by  Bozza \cite{B.V.1.}. At first, we consider two new variables $z$ and $y$ as

 \begin{equation}\label{51}
  y =1-\frac{r_0}{ r}
      \end{equation}
 Using the new variables, the total azimuthal angle $\phi(r_0)$ can be written as
 \begin{equation}\label{52}
 \phi(r_0)=\int_{0}^1 F(y,r_0)dy =\int_{0}^1 \frac{2r_0}{\sqrt{G_p(y,r_0)}}dy
 \end{equation}
 where
 \begin{equation}\label{53}
   G_p(y,r_0)=\frac{R_p(z,r_0) K(z,r_0)}{G(z,r_0)}(1-y)^4
 \end{equation}
 Above equation can be expanded as a power series of $y$
  in the following form

  \begin{equation}\label{54}
G_p(y,r_0)=\Sigma^{\infty}_{i=1} b_i(r_0) y^n
     \end{equation}
     where $ b_1(r), b_2(r) $ are given by \cite{T.4.}
     \begin{equation}\label{55}
     b_1(r)=\frac{K_0 D_0 r_0}{G_0}
     \end{equation}
     \begin{equation}\label{56}
   b_2(r) = \frac{H_0(r_0)r_0}{G_0}\{D_0[(D_0-\frac{G^{\prime}_0}{G_0})r_0-3]+\frac{r_0}{2}(\frac{K^{\prime \prime}_0}{K_0}-\frac{F^{\prime \prime}_0}{F_0})
     \end{equation}

     respectively
     \begin{equation}\label{57}
     b_2(r_c)=\frac{K_c r^2_c}{2G_c}D^{\prime}_c
     \end{equation}
     where
     \begin{equation}\label{58}
     D^{\prime}_c=\frac{K^{\prime \prime}_0}{K_0}-\frac{F^{\prime \prime}_0}{F_0}
     \end{equation}
From the above discussion, we can obtain that the leading order of
the divergence of $ F(z,r_0)$ is  $y^{-1}$, which shows that the
integral $ \phi(r_0)$ divergence logarithmically for  $r_0\rightarrow r_c$.

To solve,the integral in(\ref{52}), it can be separated into a divergent $\phi_{D}(r_0)$  and a regular part $\phi_{R}(r_0)$ as
 \begin{equation}\label{59}
     \phi(r_0) =  \phi_{D}(r_0)+\phi_{R}(r_0)
 \end{equation}

    where the divergent part is define as
      \begin{equation}\label{60}
      \phi_D(r_0)=\int_{0}^1 g_D(y,r_0)dy
      \end{equation}

     \begin{equation}\label{61}
      g_D(y,r_0)= \frac{2r_0}{\sqrt{b_(r_0) y +b_2(r_0) y^2}}
  \end{equation}
 and integrating  Eq.(60), we obtain
 \begin{equation}\label{62}
  \phi_D(r_0)=\frac{4r_0}{\sqrt{b_2(r)}}\log{\frac{\sqrt{b_2(r_0)}+\sqrt{b_1(r_0)+b_2(r_0)}}{\sqrt{b_1(r_0)}}}
 \end{equation}
and the regular part is  define as
 \begin{equation}\label{63}
 \phi_{R}(r_0)=\int_{0}^1 g_{R}(y,r_0) dy
\end{equation} .

 where
  \begin{equation}\label{64}
g_{R}(y,r_0) = g(y,r_0)-g_{D}(y,r_0)
 \end{equation}
 The  angle  of deflection (for $ r_0 \sim r_c$) can be represented as
  \begin{equation}\label{65}
     \alpha(u)= -c_{1} log(\frac{u(r)}{u(r_c)}-1) +c_{2} +\mathcal{O}(u(r) -u(r_c))
     \end{equation}

   \begin{equation}\label{66}
       c_{1}=\sqrt{\frac{2G_c}{K_m[\frac{(KW)_c^{\prime\prime}}{(KW)_c}-\frac{F^{\prime\prime}_c}{F_c}]}}
   \end{equation}
   and
   \begin{equation}\label{67}
      c_{2}=-\pi + a_R + c_{1} log\{r^2_c[\frac{(KW)^{\prime\prime}_c}{(KW)_c}-\frac{F^{\prime\prime}_c}{F_c}]\}
  \end{equation}
  $ a_R=\phi_R(r_c)= \int_{0}^{1} g_R(y,r_c ) dy  $.

Here, the subscript $c$ indicates the value of the quantities at
$r=r_c$
 \begin{figure}\label{68}
 \begin{subfigure}[b]{.4\textwidth}
    \includegraphics[width=80mm]{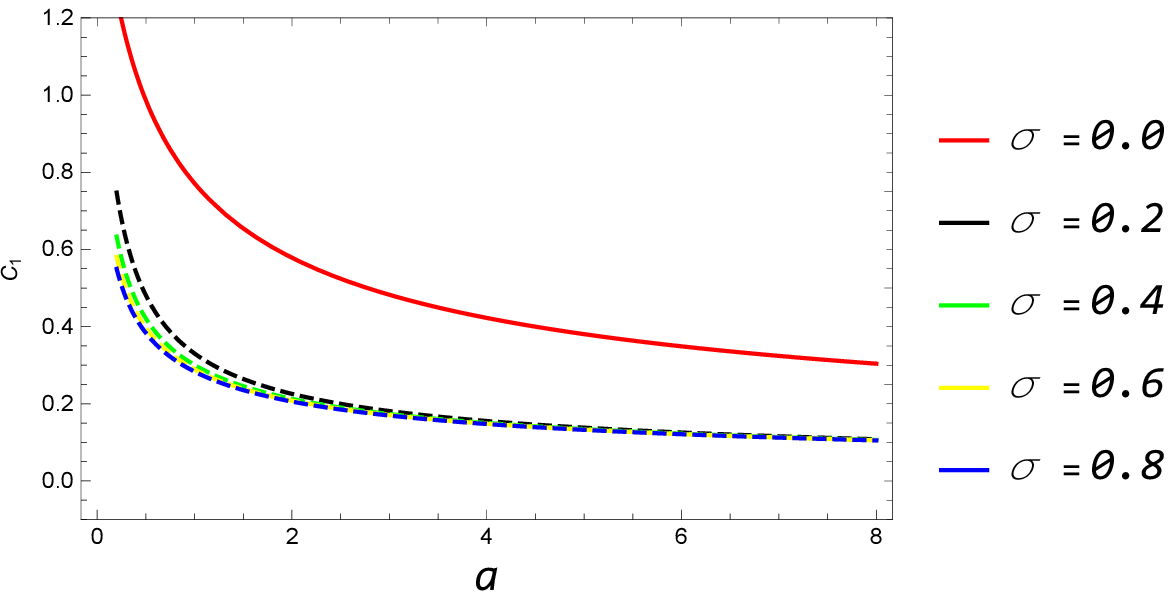}
    \caption{}
     \label{fig:f 2}
      \end{subfigure}
      \hfill
      \begin{subfigure}[b]{.4\textwidth}
    \includegraphics[width=80mm]{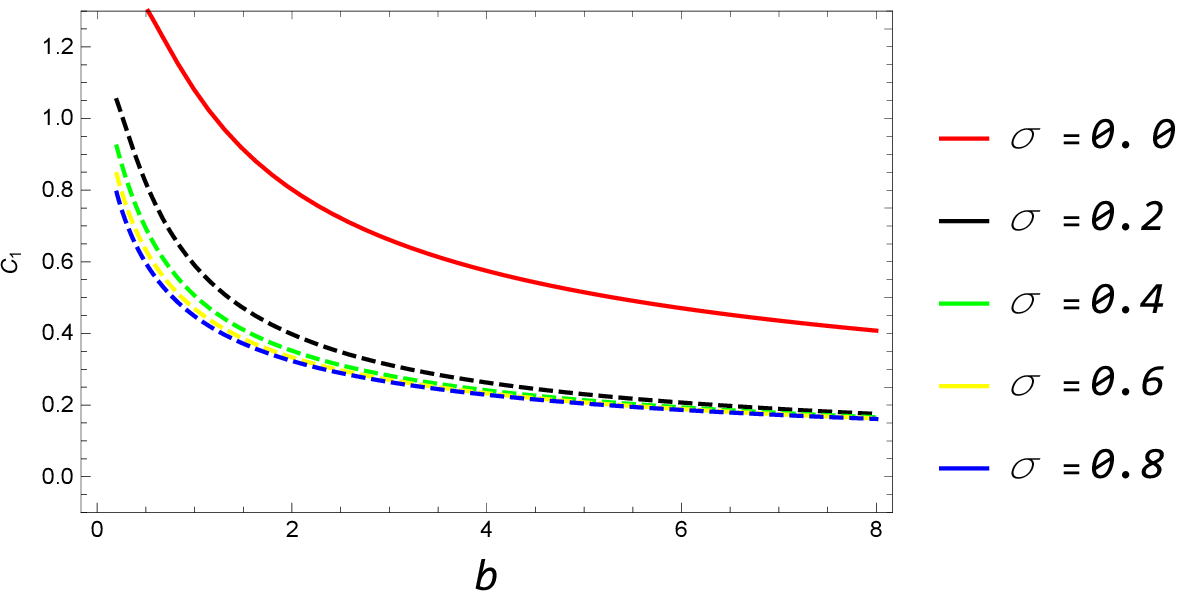}
    \caption{}
     \label{fig:f 2}

      \end{subfigure}
    \caption{The strong deflection limit coefficient $\mathit{C_1}$ as a function of  $a$  (Left panel) and  $b$  (Right panel)for the  values of $\sigma=0.0,0.2,0.4,0.4,0.6,0.8$. Parameter $\sigma =\frac{\omega_p}{\omega_{\infty}}$ represents the ratio of  plasma frequency and  photon frequency.}
    \label{fig:}
      \end{figure}

  \begin{figure}
 \begin{subfigure}[b]{.4\textwidth}
    \includegraphics[width=80mm]{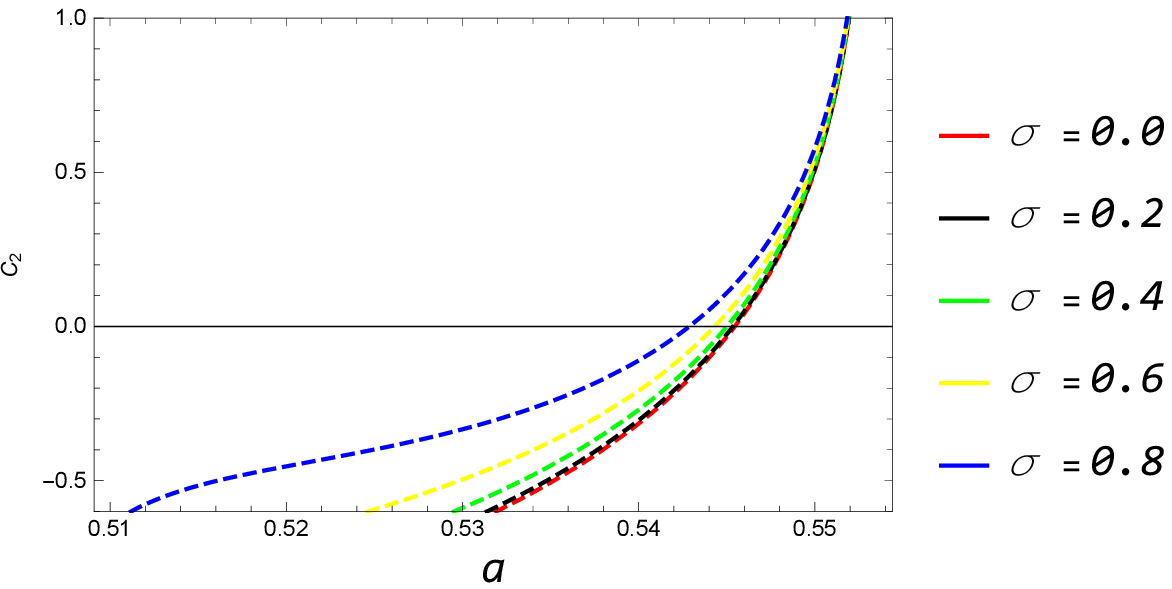}
    \caption{}
     \label{fig:f 2}
      \end{subfigure}
      \hfill
      \begin{subfigure}[b]{.4\textwidth}
    \includegraphics[width=80mm]{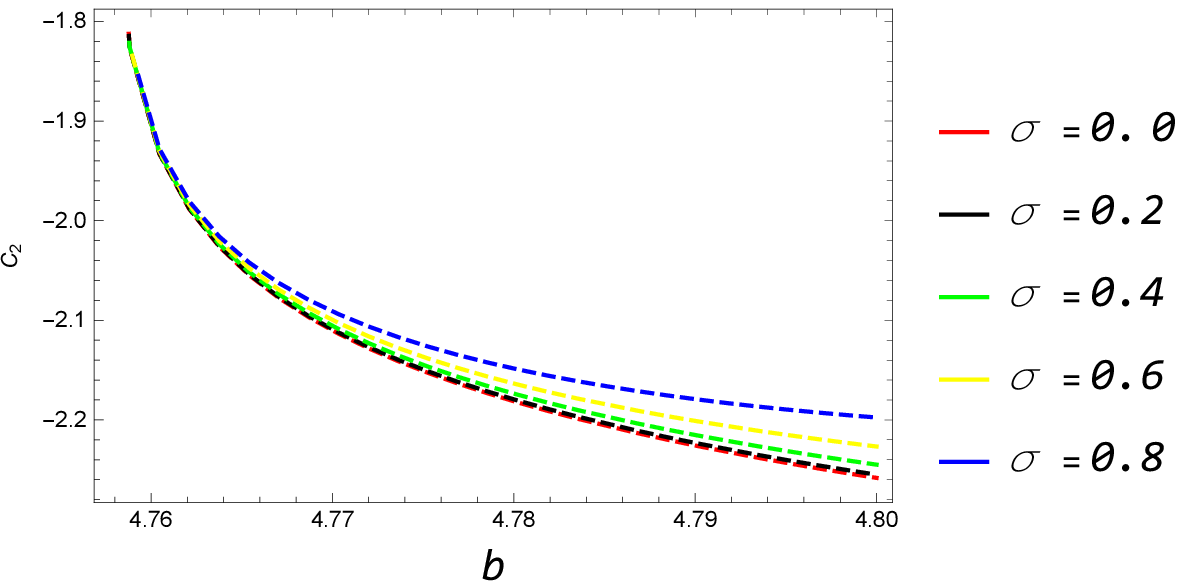}
    \caption{}
     \label{fig:f 2}

      \end{subfigure}
    \caption{The strong deflection limit coefficient $\mathit{C_2}$ as a function of  $a$  (Left panel) and  $b$  (Right panel)for the  values of $\sigma=0.0,0.2,0.4,0.4,0.6,0.8$. Parameter $\sigma =\frac{\omega_p}{\omega_{\infty}}$ represents the ratio of plasma frequency to the photon frequency.}
    \label{fig:}
      \end{figure}
 \begin{figure}
 \begin{subfigure}[b]{.4\textwidth}
    \includegraphics[width=80mm]{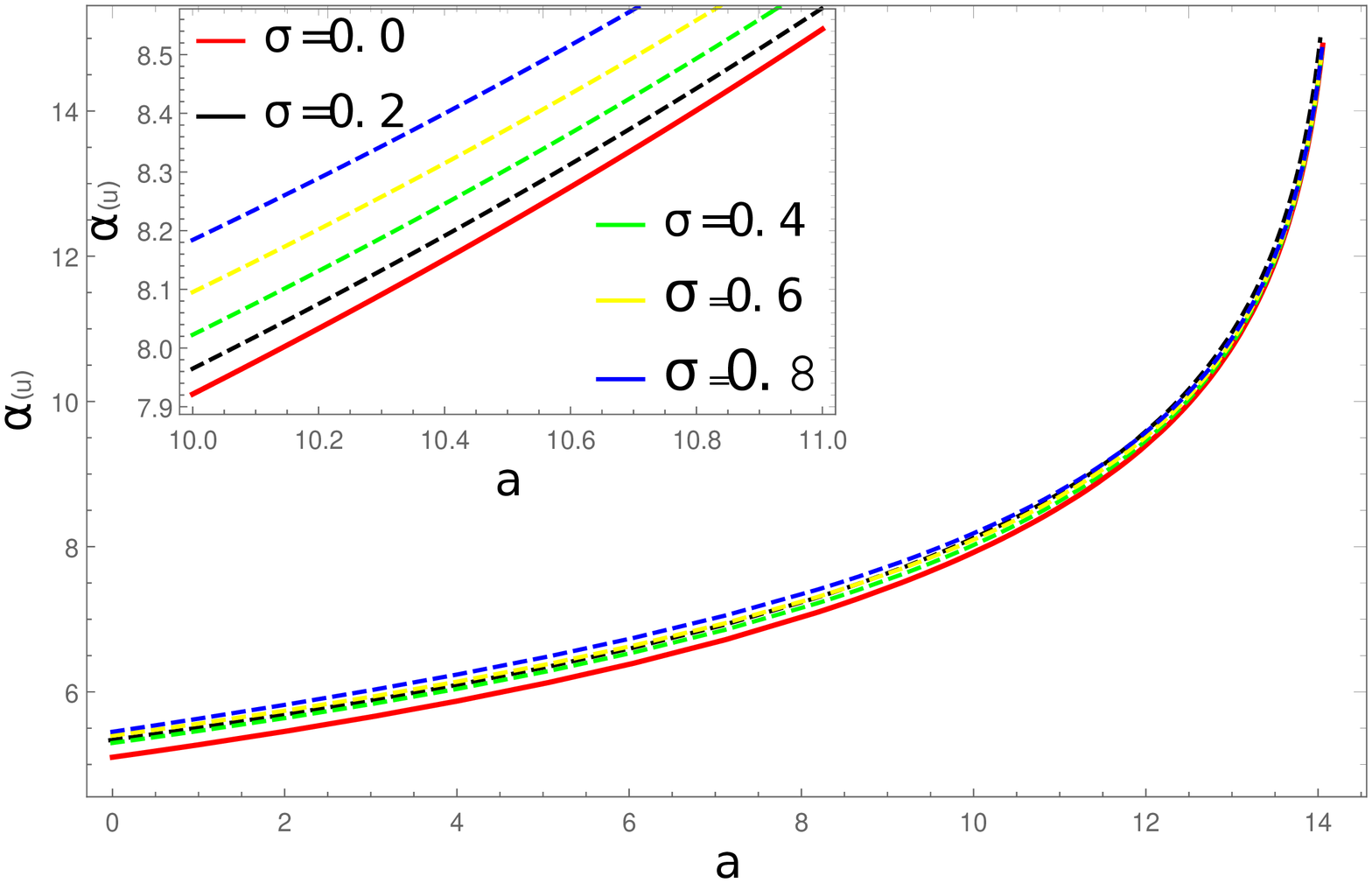}
    \caption{}
     \label{fig:f 2}
      \end{subfigure}
      \hfill
      \begin{subfigure}[b]{.4\textwidth}
    \includegraphics[width=90mm]{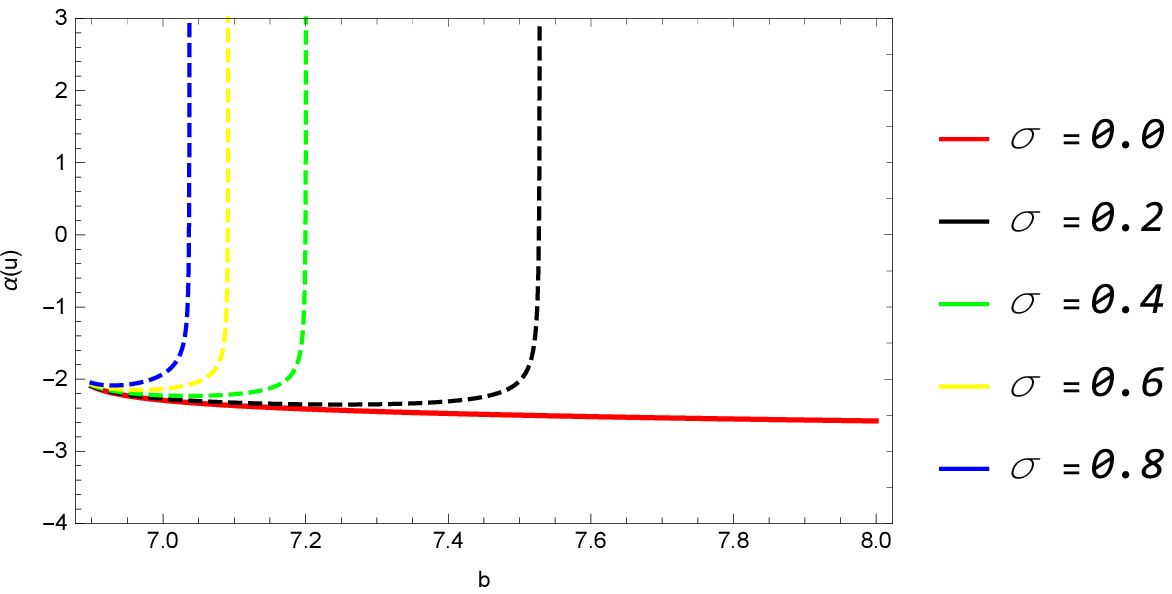}
    \caption{}
     \label{fig:f 2}

      \end{subfigure}
    \caption{The strong deflection angle $\alpha(u)$  as  a function of  $a$  (Left panel) and  $ b $  (Right panel)for the  values of $\sigma=0.0,0.2,0.4,0.4,0.6,0.8 $. Parameter $\sigma =\frac{\omega_p}{\omega_{\infty}}$ represents the ratio of  plasma frequency and  photon frequency.}
    \label{fig:}
      \end{figure}
The strong  deflection limit  coefficients $ c_1$ and $ c_2$  with
the parameter a (left panel) as well as the parameter b (right
panel) are illustrated in Figs. 9 and 10. In Fig.9 we can see that
for the fixed value of the parameter $\sigma$ ,the strong
deflection limit coefficient  $c_1$ decreases with the parameter a
(left panel) as well as the parameter b (right panel). We also can
see that the strong deflection limit coefficient  $c_1$ in a
homogeneous plasma medium is smaller than  the vacuum case.In
Fig.10 we can see that  for the fixed value of the parameter
$\sigma$ ,the strong deflection limit coefficient  $c_2$ increases
with the parameter a (left panel) while decreases with the
parameter b (right panel). We also can see that the strong
deflection limit coefficient  $c_2$ in a homogeneous plasma medium
is larger than  the vacuum case. Using the strong  deflection
limit  coefficients $ c_1$ and $ c_2$ , we can calculate the
strong deflection angle $\alpha(u)$. For the different values of
the parameter $\sigma$  , behaviour the strong deflection angle
$\alpha(u)$ with   the parameter a (left panel) as well as the
parameter b (right panel) is displayed in Fig.11.In this figure,
it is easy to obtain that  for the fixed value of the parameter
$\sigma$ ,the strong deflection angle $\alpha(u)$increases with
the parameter a (left panel) as well as the parameter b (right
panel). We also can see that the strong deflection angle
$\alpha(u)$ in a homogeneous plasma medium is larger than  that of
vacuum medium.

\section{Observables and Relativistic Images of Strong gravitational lensing}
In this section, we study about the observables quantities of
strong GL by Van der Waals BH with  the
effects of homogeneous plasma. Setting the BH at origin,
the angle between the source and optical axis  is denoted by
$\beta$. Here, we are only interested in the case in which  the
lens, source and observer are almost alignment i.e. $\beta \simeq
0$.From the lens geometry in ref.\cite{B.V.1.,B.V.2.}, the lens
equation  can be expressed in the strong field as
\begin{equation}\label{68}
\beta=(\frac{D_{ls}-D_{ol}}{D_{ls}})\theta-\alpha(\theta)mod(2\pi)
\end{equation}
where $D_{ls}$ and $D_{ol}$  is the  distance between lens and
source and distance, and  the distance between  observer and lens
respectively. The angle $\theta=\frac{u_c}{D_{ol}}$ represents the
angular image position  with respect to the optical axis.

Following Ref.\cite{B.V.2.,B.4.,B.6.}, the angular position
between the lens and the  $n^{th}$ relativistic image can be
expressed as
\begin{equation}\label{70}
\theta_{n}=\theta^0_{n}(1-\frac{u_{c}e_n(D_{ol}+D_{ls})}{\bar{a}D_{ol}D_{ls}})
\end{equation}

where
\begin{equation}\label{71}
\theta^0_{n}=\frac{u_{c}}{D_{ol}}(1+e_n),
\end{equation}
and
\begin{equation}\label{72}
e_n=e^{\frac{c_2+|\beta|-2n\pi }{c_1}}
\end{equation}

As in Ref.\cite{B.V.2.,B.4.,B.6.}, we consider a simple case in
which only the outer most image i.e., the first image $\theta_{1}$
is treated as a single image and  all the others images are
packed together at $\theta_{\infty}$.
With the help of equations (\ref{70}) and (\ref{71}),  one can obtain  the  angular position of  set of images  $\theta_{\infty}$  related  to the minimum impact parameter $u_c$ as

\begin{equation}\label{73}
\theta_{\infty}=\frac{u_c}{D_{ol}}
\end{equation}

The observable quantity $S$,
the angular image  separation  between the first image $\theta_{1}$ and the others images can be  obtained as

\begin{equation}\label{74}
s=\theta_1 -\theta_{\infty}=\theta_{\infty}e^{\frac{c_2-2n\pi }{c_1}}
\end{equation}
 and the  observable quantity  $r_{mag}$, the ratio of the flux between the first image  $\theta_{1}$ and the others images as
\begin{equation}\label{75}
r_{mag}=exp(\frac{2\pi}{c_1})
\end{equation}

If the observables quantities  $\theta_{\infty}$,s,$r_{mag} $ are
available from the observations,  the strong deflection limit
coefficients $c_1$, $c_2$ and the impact parameter $u_c$ can be
calculate easily and then compared to theoretical values.

  \begin{figure}
 \begin{subfigure}[b]{.4\textwidth}
    \includegraphics[width=80mm]{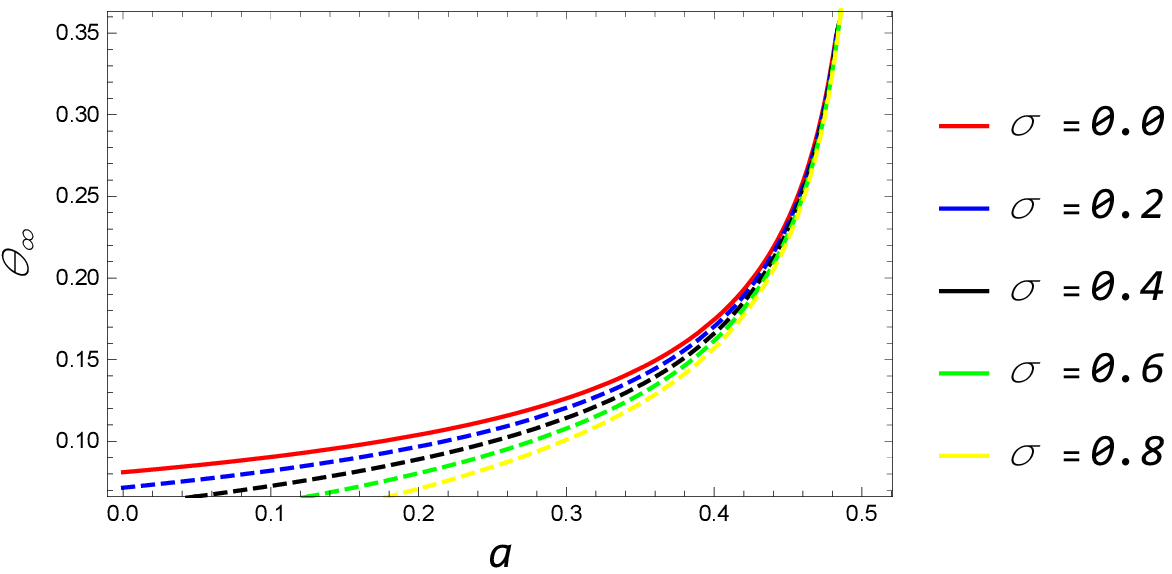}
    \caption{}
     \label{fig:f 2}
      \end{subfigure}
      \hfill
      \begin{subfigure}[b]{.4\textwidth}
    \includegraphics[width=80mm]{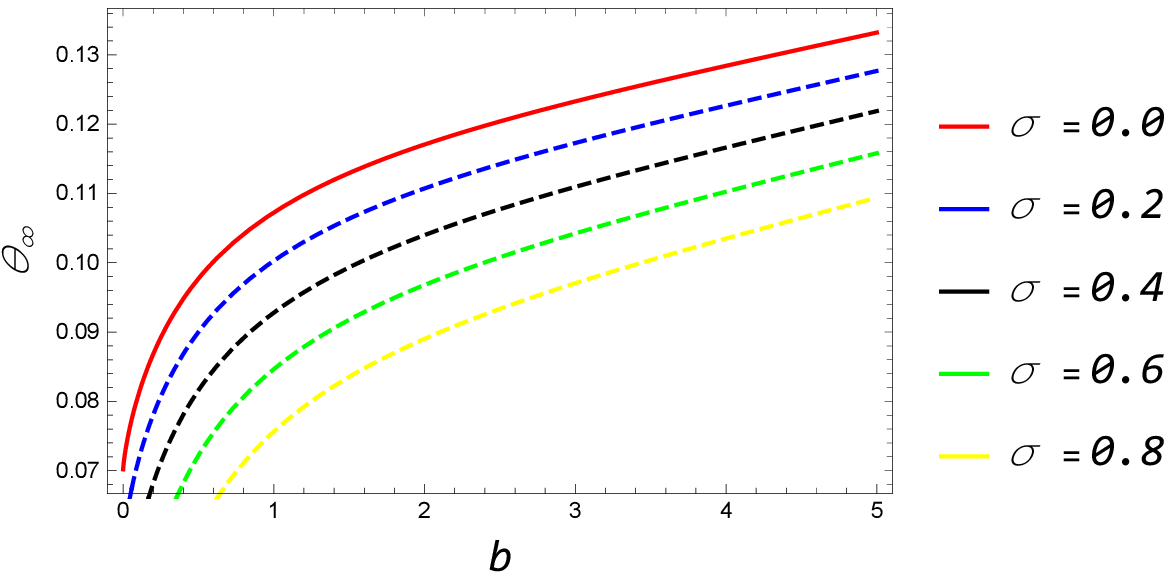}
    \caption{}
     \label{fig:f 2}

      \end{subfigure}
    \caption{The angular  position  of image $\theta_{\infty}$ as a function of  $a$  (Left panel) and  $b$  (Right panel)for the  values of $\sigma=0.0,0.2,0.4,0.4,0.6,0.8$. Parameter $\sigma =\frac{\omega_p}{\omega_{\infty}}$ represents the ratio of  plasma frequency and  photon frequency.}
    \label{fig:}
      \end{figure}

  \begin{figure}
 \begin{subfigure}[b]{.4\textwidth}
    \includegraphics[width=80mm]{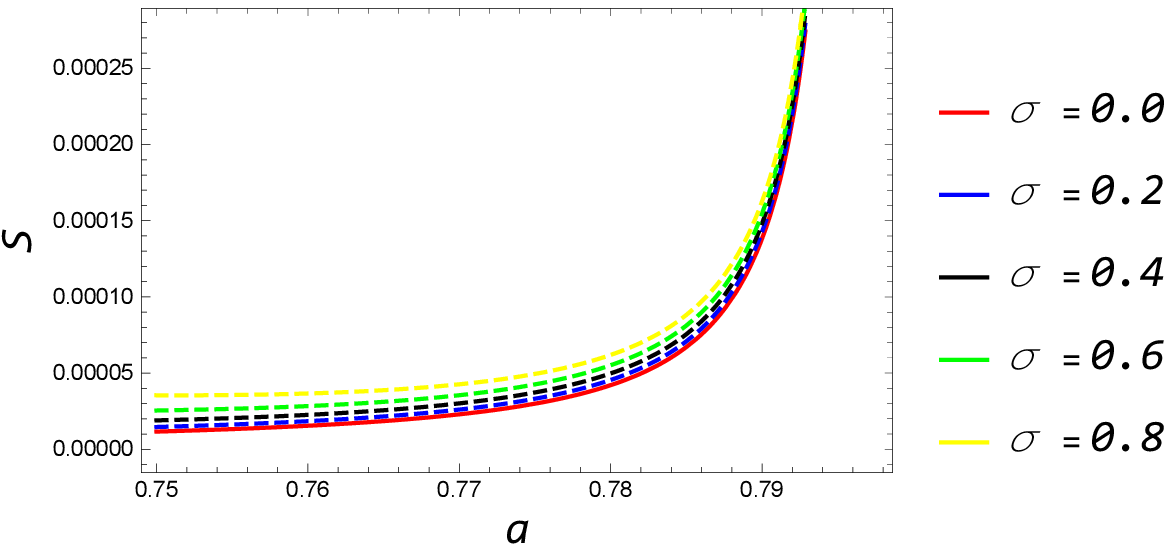}
    \caption{}
     \label{fig:f 2}
      \end{subfigure}
      \hfill
      \begin{subfigure}[b]{.4\textwidth}
    \includegraphics[width=80mm]{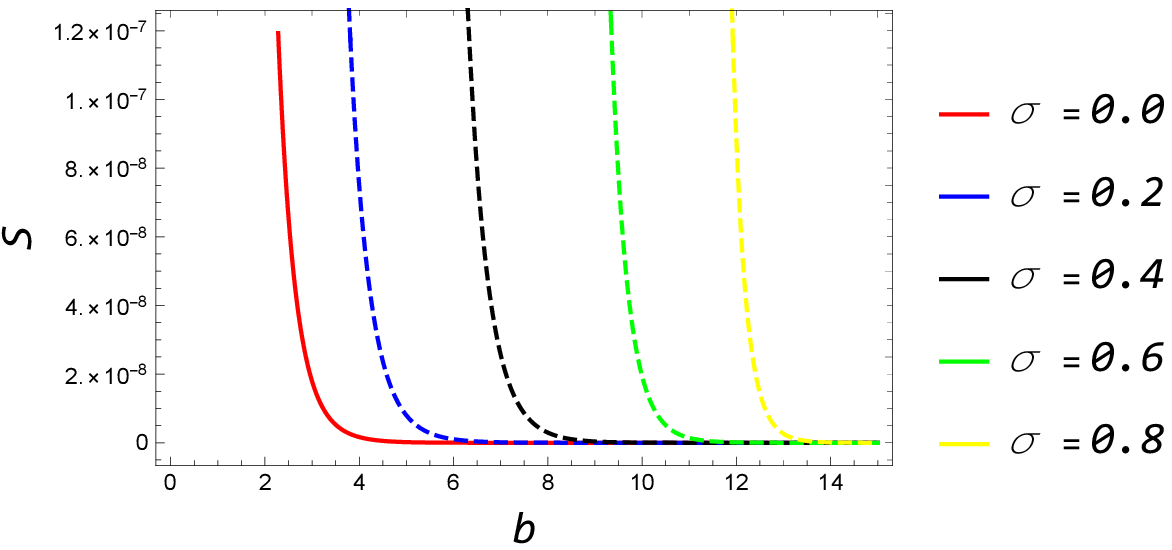}
    \caption{}
     \label{fig:f 2}

      \end{subfigure}
    \caption{The angular separation  of the images $s$ as a function of  $a$  (Left panel) and  $b$  (Right panel)for the  values of $\sigma=0.0,0.2,0.4,0.4,0.6,0.8$. Parameter $\sigma =\frac{\omega_p}{\omega_{\infty}}$ represents the  ratio of  plasma frequency and  photon frequency.}
    \label{fig:}
      \end{figure}

  \begin{figure}
 \begin{subfigure}[b]{.4\textwidth}
    \includegraphics[width=80mm]{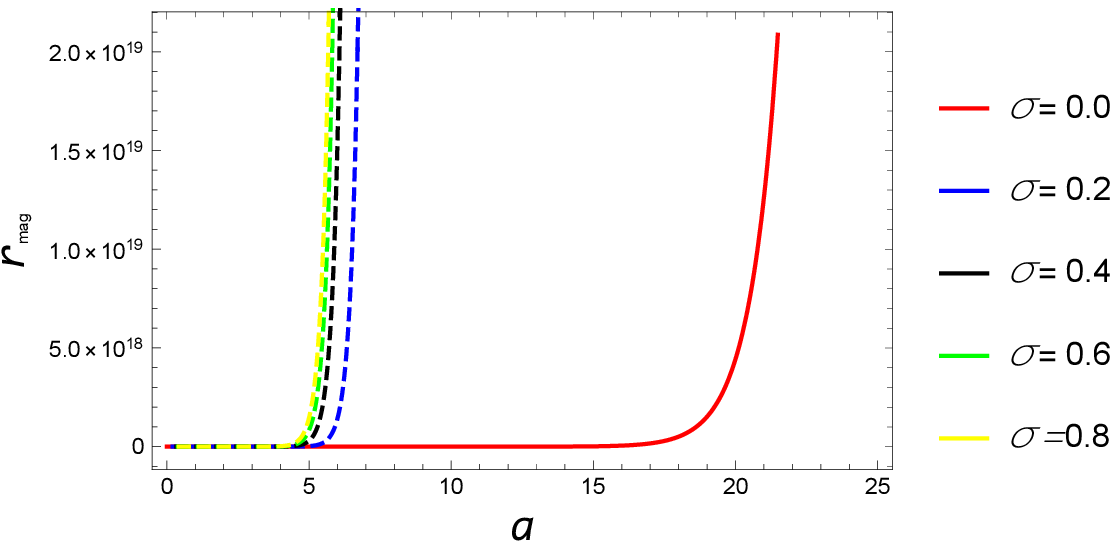}
    \caption{}
     \label{fig:f 2}
      \end{subfigure}
      \hfill
      \begin{subfigure}[b]{.4\textwidth}
    \includegraphics[width=80mm]{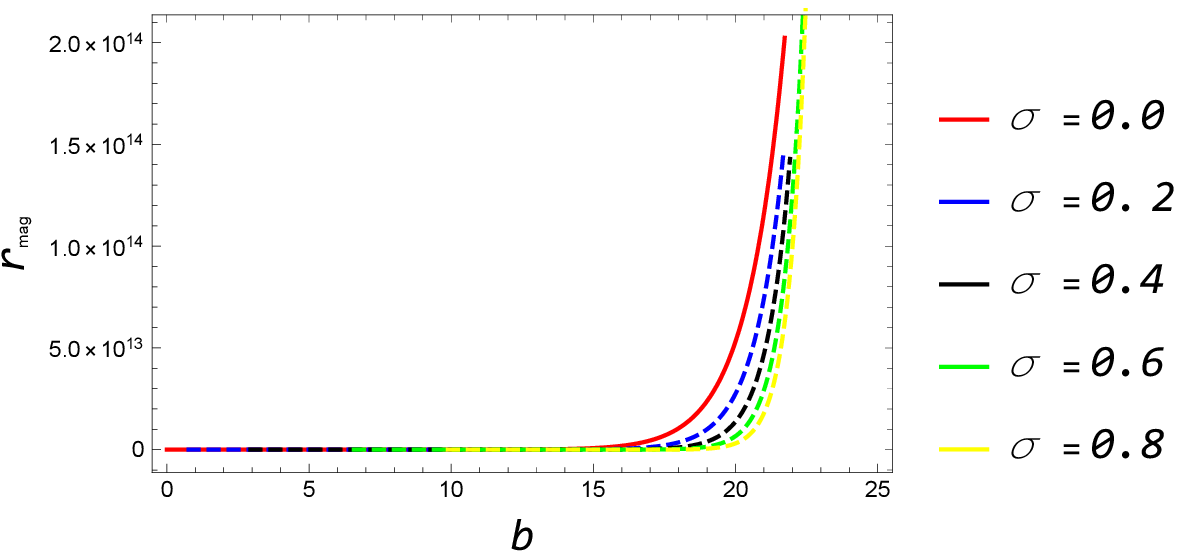}
    \caption{}
     \label{fig:f 2}

      \end{subfigure}
    \caption{The relative magnification   $r_{mag}$ as a function of  $a$  (Left panel) and  $b$  (Right panel)for the  values of $\sigma=0.0,0.2,0.4,0.4,0.6,0.8$. Parameter $\sigma =\frac{\omega_p}{\omega_{\infty}}$ represents the ratio of  plasma frequency and  photon frequency.}
    \label{fig:}
      \end{figure}

We plot the strong observable quantities $\theta_{\infty}$, $S$
and $r_{mag}$ in a realistic scenario of the BH, such as $SgrA^*$
having mass $M=6.5 \times 10^6 M_{\odot}$ and $D_{ol}=16.8Mpc$
\cite{B.S.S.} are displayed in Figs.(12),(13),and (14)by assuming
$D_{ls}=1kpc$. We can see that for the fixed value of the
parameter $\sigma$, the strong observable quantities
$\theta_{\infty}$ and $r_{mag}$ increases with the parameter $a$
(left panel) as well as the parameter $b$ (right panel),
respectively in Fig.13 and Fig.14. We also can see that strong
observable quantity $\theta_{\infty}$  in a homogeneous plasma
medium is smaller than that of vacuum medium with respect to the
parameter $a$ as well as the parameter $b$. But the strong
observable quantities $r_{mag}$ in a homogeneous plasma medium is
smaller than the vacuum medium with respect to the parameter $a$
while larger than the vacuum medium with respect to the parameter
$b$. In Fig.13, we can see that with the fixed value of the
parameter $\sigma$, the angular separation S increases with the
parameter a (left panel) and decreases with the parameter $b$
(right panel). In this figure, we can also see that the angular
separation $S$ in a homogeneous plasma medium is smaller than that
of a vacuum medium with respect to the parameter $a$ as well as
the parameter $b$.

\section{Discussions and Conclusions}
In this work, we have discussed the Shadows and strong
gravitational lensing (GL) by Van der Waals BH in the presence of
a homogeneous plasma medium. At first, we analyzed the horizon
structure of the Van der Waals BH. Next, we have analytically
calculated the first order null geodesic equation by the
Hamiltonian-Jacobi separation method. As an application of null
geodesic equations, one can determine BH shape as well as shadow
shape. With the help of these geodesics equations, we calculate
and discuss the shadows of the Van der Waals BH in the absence of
plasma medium as well as the presence of plasma medium,
respectively. In this work, we graphically calculate the radius of
the photon sphere with the different values of parameters $a$, $b$
and with the different values of parameters $\sigma$,
respectively. We observe that for the fixed value of the parameter
$a$, radius of shadows of the BH decreases while the parameter $b$
increases in the absence of a plasma medium. In the presence of a
homogeneous plasma medium, we have calculated the circular radius
of the photon sphere for the different values of the parameter
$\sigma$ (the Parameter $\sigma =\frac{\omega_p}{\omega_{\infty}}$
represents the ratio of plasma frequency and photon frequency)
graphically.

We have also seen that the radius of shadows of the BH in a homogeneous plasma medium decreases while the parameter $\sigma$ increases and the radius of shadows in a homogeneous plasma medium is larger than the vacuum medium. Next, we studied the strong GL with the effects of a homogeneous plasma medium. We have studied the effects of VdW parameters $a$, $b$ and the parameter $\sigma$ (the ratio of plasma frequency to the photon frequency ) on the strong deflection angle $\alpha(u)$ and strong lensing observables $\theta_{\infty}$,$S$, $r_{mag}$ in a homogeneous plasma medium and compared to the vacuum medium. We find that for the fixed value of $\sigma$, the strong deflection angle increases with the parameters $a$  as well as $b$. We also see that the deflection angle in the strong field with the presence of a homogeneous plasma medium is larger than the vacuum cases. We have studied the observables quantities angular position $\theta_{\infty}$, separation $S$ and magnification $r_{mag}$ by taking the example of super massive BH $SgrA^*$ in the strong field limit with the effects of homogeneous plasma. We have seen that strong observable quantities $\theta_{\infty}$ and $S$  in a homogeneous plasma medium are smaller than that of a vacuum medium with respect to the parameter $a$ as well as the parameter $b$ and the strong observable quantities $r_{mag}$ in a homogeneous plasma medium is smaller than the vacuum medium with respect to the parameter $a$ while larger than the vacuum medium with respect to the parameter $b$.\\\\

\textbf{ Acknowledgement:} NUM is thankful to CSIR, Govt. of
India for providing Senior Research Fellowship (No. 08/003(0141))/2020-EMR-I).\\

\end{document}